\documentclass[reqno,oneside]{amsart}

\usepackage{graphicx,cite,mathptmx,times}

\DeclareMathOperator{\dist}{dist}
\DeclareMathOperator{\Arg}{Arg}
\DeclareMathOperator{\sgrad}{sgrad}
\DeclareMathOperator{\spec}{spec}
\DeclareMathOperator{\hol}{hol}
\DeclareMathOperator{\vp}{v.~p.}
\DeclareMathOperator{\Ind}{ind}

\begin{document}

{

\noindent\LARGE\bf On semiclassical dispersion relations of Harper-like operators

}

\Large

\vskip 10mm

\leftskip 30mm

\noindent Konstantin Pankrashkin

\vskip 5mm

\normalsize

\noindent Institut f\"ur Mathematik, Humboldt-Universit\"at
zu Berlin\\Rudower Chaussee~25, Berlin 12489 Germany\\[\medskipamount]
E-mail: \texttt{const@mathematik.hu-berlin.de}

\vskip 15mm

\noindent\textbf{Abstract.}
We describe some semiclassical spectral properties of
Harper-like operators,
i.~e. of one-dimensional quantum Hamiltonians periodic
in both momentum and position. The spectral region corresponding
to the separatrices of the classical Hamiltonian is studied
for the case of integer flux. We derive asymptotic formula
for the dispersion relations, the width of bands and gaps,
and show how geometric characteristics and the absence
of symmetries of the Hamiltonian influences the form 
of the energy bands.

\leftskip=0mm

\bigskip

\section{Introduction}

In the present work we are going to describe
certain asymptotic spectral properties of the Harper-like
operators. Such operators appear as follows.
Let $H(p,x)$ be a real-valued real-analytic one-dimensional
classical Hamiltonian periodic in both momentum and position:
\[
H(p+2\pi,x)\equiv H(p,x+2\pi)\equiv H(p,x),
\quad p,x\in\mathbb{R}.
\]
The operator $\Hat{H}_h$ obtained from $H$ through
the Weyl quantization, 
\[
\big(\Hat{H}_h f\big)(x)=
\frac{1}{2\pi h}\int_{\mathbb{R}}\int_{\mathbb{R}}
e^{ip(x-y)/h}
H\big(p,\frac{x+y}{2}\big) f(y)\,dp\,dy,
\]
will be called \emph{the Harper-like operator associated with $H$}.
In one of the simplest cases, when $H(p,x)=2\cos p+2\alpha\cos x$,
one has $\big(\Hat{H}_h f\big)(x)=
f(x+h)+f(x-h)+2\alpha\cos x\ f(x)$,
i.~e. $\Hat{H}_h$ is the Harper operator on the real line.

Such operators appear in the study of the motion of electrons in a two-dimensional periodic potential
subjected to a uniform perpendicular magnetic field~\cite{harper,azbel}, and
the parameter $h$ may be expressed through the parameters
of the system differently in the cases of strong and weak magnetic fields~\cite{langbein}.
The periodic magnetic systems show a rich spectral structure depending on $h$,
which in general expresses a relationship between the magnetic and electric fields;
for example,
in the strong magnetic field one can put $h=(l_M/L)^2$, where $l_M$ is the magnetic length,
and $L$ is the characteristic size of the period lattice~\cite{bdgp}.
The Hamiltonian $\Hat H_h$ commutes with operators $T_j$, $j=1,2$, defined by
\[
(T_1 f)(x)=e^{-2\pi i x/h} f(x),\quad
(T_2 f)(x)=f(x+2\pi);
\]
they obey the equality $T_1 T_2=\exp (4\pi^2 i/h) T_2 T_1$
and commute iff $\eta:=2\pi/h\in\mathbb{Z}$; this number $\eta$ is usually referred to
as the number of magnetic flux quanta through an elementary cell or simply \emph{flux}.
In the case of integer $\eta$ one can apply the usual Bloch theory
and show that the spectrum consists of $\eta$ bands~\cite{faure},
so that each band is the value set of the corresponding
dispersion relation $E(\mathbf{k},h)$, where
$\mathbf{k}=(k_1,k_2)\in[-\pi,\pi)\times[-\pi,\pi)$
is the (vector) quasimomentum, and for any $\mathbf{k}$ there exists
a (generalized) eigenfunction $\Psi(x,\mathbf{k},h)$ of $\Hat H_h$,
$\Hat H_h\Psi(x,\mathbf{k},h)=E(\mathbf{k},h)\Psi(x,\mathbf{k},h)$,
satisfying the Bloch-periodicity conditions
\begin{equation}
              \label{eq-bloch}
T_j \Psi(x,\mathbf{k},h) =e^{i k_j}\Psi(x,\mathbf{k},h),\quad j=1,2.
\end{equation}
The situation with non-integer but rational $\eta=N/M$ ($N$ and $M$ are mutually prime integers)
can be reduced to the previous case by enlarging the unit
cell of the period lattice $(2\pi,2\pi)\rightarrow(2\pi M,2\pi)$.
For irrational $\eta$ the spectrum is rather complicated
and can include parts of Cantor structure as was predicted in~\cite{azbel,hof}
and then analytically justified for the Harper operator in~\cite{wilk1,helf}.

We will study the asymptotics of the spectrum for $\Hat H_h$ for large integer $\eta$, which
corresponds to the semiclassical limit ($h\to 0$). Spectral bands in different parts of the spectrum
have then different asymptotic behavior~\cite{faure,bdp}. To illustrate this, we consider first the
Harper operator. To be more definite, assume $0<\alpha<1$, then, for any $\delta>0$, the bands lying
below $(-1+\alpha-\delta)$ and over $(1-\alpha+\delta)$ correspond to finite classical motion
and have the width $o(h^\infty)$, while the
bands inside the segment $[-1+\alpha+\delta,1-\alpha-\delta]$ correspond to open classical
trajectories and have the width $\sim h$ with $o(h^\infty)$-gaps between them~\cite{wt}.
These estimates are non-uniform with respect to $\delta$, and they do not describe the bandwidth
asymptotics near the critical points $\pm(1-\alpha)$, where the eigenfunctions undergo a transition
from localized to extended behavior~\cite{aa}. For Hamiltonians of a more general form the spectral
structure is suitably described by the Bohr-Sommerfeld quantization rule and can be illustrated with
the help of the so-called Reeb graph technique~\cite{bdp,faure}, so that critical values of $H$
divide the spectrum into parts with different asymptotic behavior (we illustrate this
in section~\ref{sec2}), and the transitions between these
parts semiclassically correspond to the separatrices of the classical Hamiltonian.

This transient region was earlier studied from different point
of view~\cite{helf,lw,wilk1} for the Harper operator and some closed to it Hamiltonians
with high-order symmetry. From the other side, the absence
of symmetries appears to be the generic situation, which influences
the structure of spectral gaps and dispersion relations, which are very different
from the ideal model of the Harper operator; this leads
to new phenomena in the transport properties of the periodic magnetic systems,
see the review~\cite{dem}.
(This covers not only the Harper-like operators, but also the periodic
Schr\"odinger operators, where symmetry properties are in connection
with the absolute continuity of the spectrum~\cite{fil}.)
Our aim here is to provide a uniform approach to studying the band- and gapwidth transition
and construction of the dispersion relations. In particular, we are interested in the way
how the shape of the classical trajectories influences the shape of the energy bands.
We study in details only some basic cases (sections~\ref{sec4} and~\ref{sec5}), but even these simple
examples show a rich structure of the dispersion relations; in particular,
it turns out that the shape of the energy bands as well as the ratio bandwidth/gapwidth
in non-symmetric cases is very sensitive to the parameters of the system;
this is discussed in section~\ref{sec6}.

It is worthwhile to emphasize that even the spectral problem for a single two-dimensional
periodic system can lead to a number of Harper-like operators with essentially different
spectral properties. We illustrate this with the example of
the two-dimensional Landau operator with a periodic electric potential $v$; in the case
of strong magnetic field the Schr\"odinger operator is
\begin{equation}
             \label{eq-landau}
\Hat L:=\frac{1}{2}\Big(-ih\frac{\partial}{\partial x_1}+x_2\Big)^2
-\frac{h^2}{2} \frac{\partial^2}{\partial x_2}+\varepsilon v(x_1,x_2),
\end{equation}
with small $h$ and $\varepsilon$~\cite{bdgp}. The spectral problem for $\Hat L$
is reduced to a series of spectral problems for the Harper-like operators
associated with the classical Hamiltonians
\begin{equation}
               \label{eq-land-n}
L_n(Y_1,Y_2)=(n+\frac{1}{2})h+\varepsilon J_0(\sqrt{-(2n+1)h\,\Delta_Y})v(Y_1,Y_2)+O(\varepsilon^2),
\qquad n\in\mathbb{Z}_+,
\end{equation}
where $J_0$ is the Bessel function of order zero, $\Delta_Y=\partial^2/\partial Y_1^2+
\partial^2/\partial Y_2^2$ (we explain this reduction in Appendix), and $(Y_1,Y_2)$ are canonically
conjugate. The Hamiltonians $L_n$ have, generally speaking, trajectories of different kind for
different $n$, which results in a difference between their spectra.

\section{Regular Bohr-Sommerfeld rules and
the asymptotics of the dispersion relations}\label{sec2}

In this section, we recall some simple constructions which are useful for estimating the
dispersion relations~\cite{bdp,faure}.

\subsection{Reeb graph}

\begin{figure}
\begin{minipage}{90mm}\centering
\includegraphics[height=45mm]{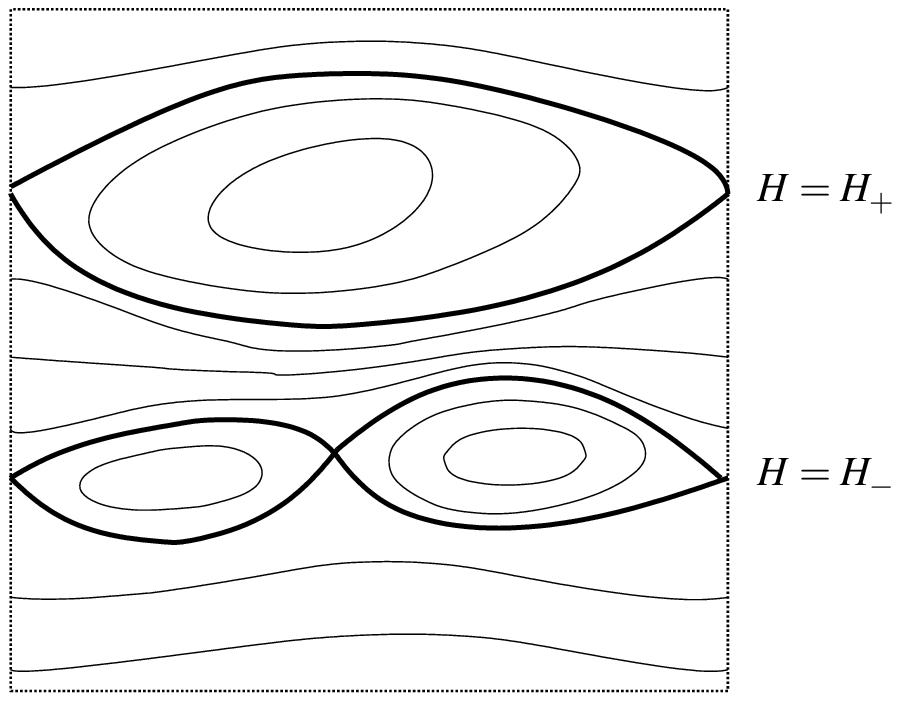}\\
(a)
\end{minipage}
\begin{minipage}{45mm}\centering
\includegraphics[height=45mm]{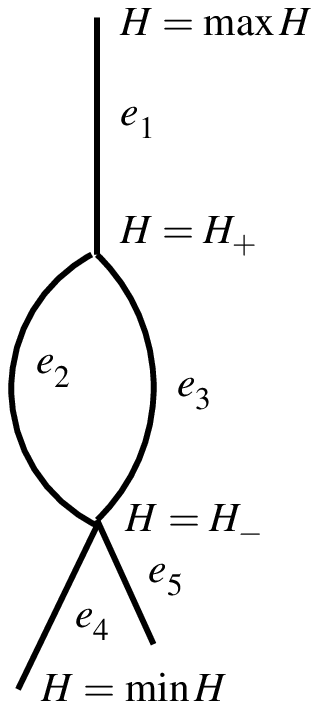}\\
(b)
\end{minipage}
\caption{Space of trajectories as the Reeb graph. (a) Level curves
of the Hamiltonian $H$ in the unit cell. (b) The corresponding
Reeb graph. Separatrices are bold and
correspond
to the branching points of the Reeb graph.}
         \label{fig0}
\end{figure}
Throughout the paper we assume that all the critical points of $H$ are non-degenerate.
Define on the torus $\mathbb{T}^2_{px}:=\mathbb{R}^2_{px}/(2\pi,2\pi)$, which will be called the
\emph{reduced phase space}, an equivalence relation $\sim$ by $x_1\sim x_2 \Longleftrightarrow \{x_1$ and
$x_2$ lie in a connected component of a level set of $H\}$, then the set $G:=\mathbb{T}_{px}^2/\sim$ is a
certain finite graph called the \emph{Reeb graph} of $H$. The end points of the graph correspond to
extremum points of the Hamiltonian while the branching points correspond to saddle points and separatrices
(see illustration in figure~\ref{fig0}). It is natural to distinguish between edges corresponding to open
trajectories on $\mathbb{R}^2_{px}$ and, respectively, to non-contractible trajectories on
$\mathbb{T}^2_{px}$ (these edges of the Reeb graph will be referred to as \emph{edges of infinite motion})
and to closed ones on the plane and contractible ones on the torus (\emph{edges of finite motion}).
This graph provides a good illustration for the structure of the trajectory space of $H$
as well as for the semiclassical spectral asymptotics.

\begin{figure}
\begin{minipage}{50mm}\centering
\includegraphics[height=45mm]{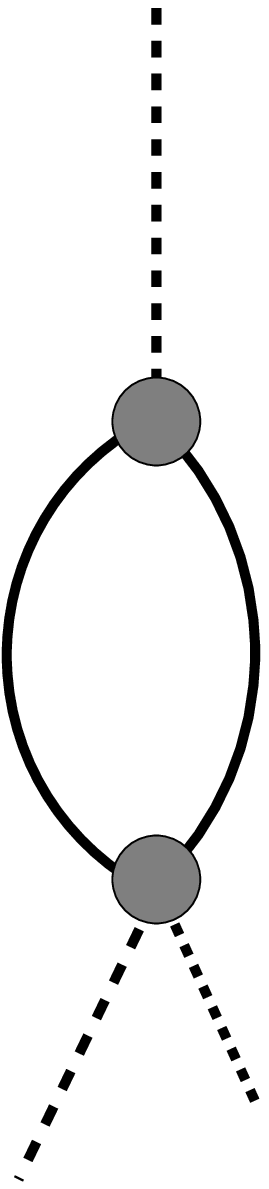}\\
(a)
\end{minipage}
\begin{minipage}{95mm}\centering
\includegraphics[height=45mm]{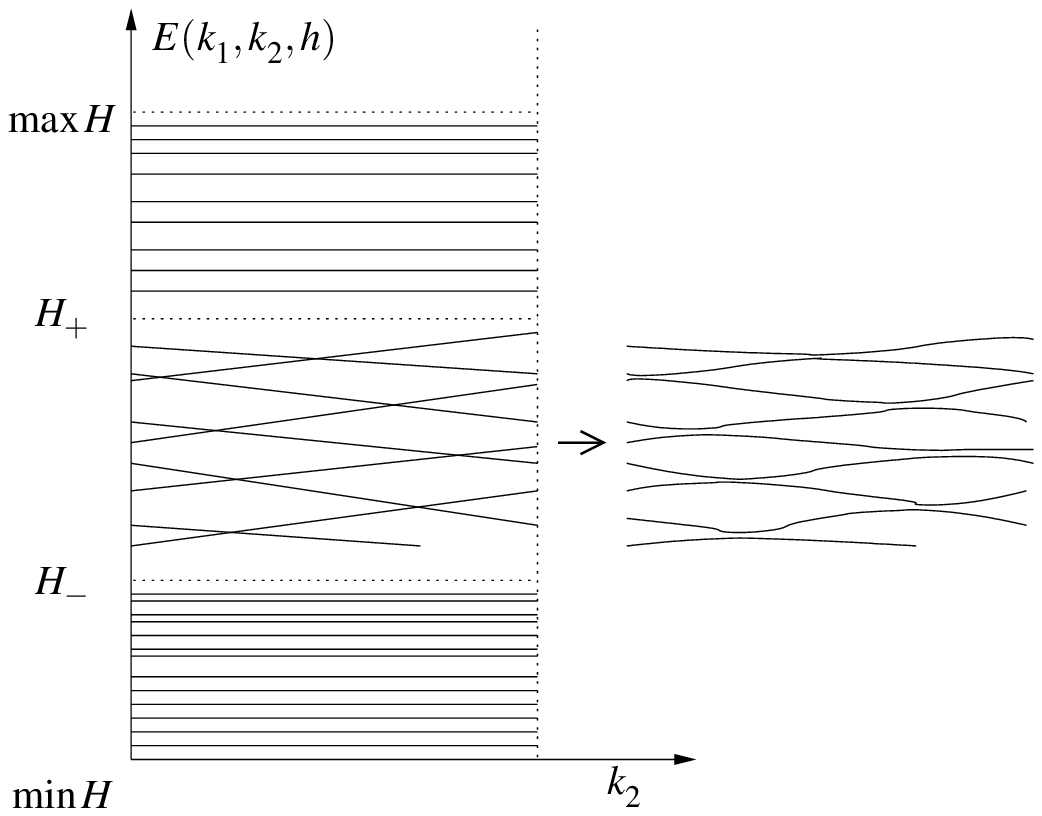}\\
(b)
\end{minipage}
\caption{Spectrum obtained with the help of the regular Bohr-Sommerfeld rules.
(a) Point of the Reeb graph, which correspond to the trajectories satisfying
the quantization condition; (b) Semiclassical dispersion relations outside the
separatrix region}
        \label{fig-quant}
\end{figure}

\subsection{Finite motion} Consider a spectral interval $\Delta=[E',E'']$ such that for each $E\in\Delta$
the level set $H=E$ on the reduced phase space $\mathbb{T}^2_{px}$ consists of a single closed
trajectory. In other word, the corresponding  part of the Reeb graph must be an edge of finite motion; in
the example shown in figure~\ref{fig0} the interval $[H_++\delta,\max H]$, $\delta>0$, satisfies this
condition; the corresponding edge is $e_1$. The regular Bohr-Sommerfeld rules select from the family of all
these closed trajectories a discrete family of trajectories satisfying the quantization condition
\[
\frac{1}{2\pi h}\oint p\,dx-\frac{1}{2}\in\mathbb{Z}.
\]
(The construction of these trajectories can be interpreted as a selection of a certain discrete subset on all the edges of finite motion of the Reeb graph,
see figure~\ref{fig-quant}a).
Each of these trajectories, $\Lambda$, implies a
quasimode $(E,\psi)$, $E\in\mathbb{R}$, $\psi(x,h)\in L^2_x(\mathbb{R})$, with $\psi$ microlocally
supported by $\Lambda$, $E=H|_{\Lambda}+O(h^2)$, and
$\big\|(\Hat{H}_h-E)\psi\big\|/\|\psi\|=o(h^\infty)$, which means that
$\dist(\spec\Hat{H}_h,E)=o(h^\infty)$.
Clearly, to each of such $E$ there corresponds a whole family of
quasimodes. Namely, if $\Lambda$ is a closed trajectory satisfying the quantization condition and
$(E,\psi)$ is the corresponding quasimode, then the trajectory $\Lambda_{\mathbf{j}}:=\Lambda+(2\pi
j_1,2\pi j_2)$, $\mathbf{j}=(j_1,j_2)\in\mathbb{Z}^2$, also satisfies the quantization condition and
produces the quasimode $(E,\psi_{\mathbf{j}})$ with \emph{the same} value $E$ and
$\psi_{\mathbf{j}}(x,h)=\exp (2\pi i j_1 x/h) \psi(x-2\pi j_2,h)\equiv T_1^{-j_1}T_2^{-j_2}\psi(x,h)$. Let us
try to satisfy the Bloch conditions by a function of the form 
\begin{equation}
       \label{bloch-ansatz-1}
\Psi(x,\mathbf{k},h)=\sum_{\mathbf{j}=(j_1,j_2)\in\mathbb{Z}^2} C_{\mathbf{j}}(\mathbf{k},h)
\psi_{\mathbf{j}}(x,h)= \sum_{\mathbf{j}=(j_1,j_2)\in\mathbb{Z}^2} C_{\mathbf{j}}(\mathbf{k},h)
T_1^{-j_1}T_2^{-j_2}\psi(x,h).
\end{equation}
Clearly, the coefficients $C_{\mathbf{j}}$ must solve a linear system:
$C_{j_1+1, j_2}=C_{j_1,j_2} e^{ik_1}$, $C_{j_1, j_2+1}=C_{j_1,j_2}
e^{ik_2}$, $j_1,j_2\in\mathbb{Z}$, therefore, $C_{\mathbf{j}}=c e^{i\langle\mathbf{j}|\mathbf{k}\rangle}$
for some constant $c$, and the corresponding Bloch quasimodes have the form \begin{equation}
\label{bloch-fun-1} \Psi(x,\mathbf{k},h)= c \sum_{\mathbf{j}=(j_1,j_2)\in\mathbb{Z}^2}
e^{i\langle\mathbf{j}|\mathbf{k}\rangle} e^{2\pi i j_1 x/h} \psi(x-2\pi j_2,h). \end{equation} Therefore,
to construct a Bloch quasimode in the form~\eqref{bloch-ansatz-1} we do not need to satisfy any
additional conditions about relationship between $E$ and $\mathbf{k}$, which
results in semiclassically constant dispersion relations.

If for each $E\in\Delta$ the level set $H$ contains several closed trajectories (and intersect several
edges of finite motion (for example, it is the interval $[\min H,H_--\delta]$ in the example of
figure~\ref{fig0}), the procedure described above is still applicable and gives
the asymptotics of the spectrum up to
$o(h^\infty)$, but, in the case of some symmetries between families of trajectories degeneracies of the
eigenvalues may occur; computation of their splitting is much more delicate~\cite{faure2,faure}.

\subsection{Infinite motion}

Now consider the spectrum in the interval
$\Delta=[E',E'']$ assuming that the corresponding region of the Reeb graph
consists of two edges of infinite motion (the interval
$[H_-+\delta,H_+-\delta]$  and the edges $e_2$
and $e_3$ in the example of figure~\ref{fig0}),
i.e. for any $E\in\Delta$ the level set $H=E$ on $\mathbb{T}^2_{px}$
consists of two non-contractible trajectories, and in $\mathbb{R}^2_{px}$
there are two families of open periodic trajectories.
Clearly, there exists
a vector $\mathbf{d}=(d_1,d_2)\in\mathbb{Z}^2$, non-divisible by
any other vector with integer-values components,
such that for any of these trajectories, $\Lambda=
\big(p=P(t),\ x=X(t)\big)$,
there exists a nonzero number $T=T(\Lambda)$ satisfying
$(P,X)(t+T)=(P,X)(t)+2\pi\mathbf{d}$ for all $t\in\mathbb{R}$.
To simplify the calculation we assume that $\mathbf{d}=(1,0)$
(otherwise one can choose new canonical coordinates on the phase
plane, such that $\mathbf{d}=(1,0)$ in these new coordinates).
Obviously, there are two families of trajectories
corresponding to different edges of the Reeb graph:
for the first family, the number $T$
can be chosen positive, while for the another one it must be negative;
we denote these families by $e^+$ and $e^-$ respectively.

Each trajectory $\Lambda^\pm=\big(p=P(t),x=X(t)\big)\in e^\pm$ implies
a quasimode $(E^\pm,\psi^\pm)$, $E^\pm=H|_{\Lambda^\pm}+O(h^2)$,
satisfying
$T_2\psi^\pm(x,h)\equiv\psi^\pm(x+2\pi,h)=\exp(iS^\pm(\Lambda^\pm)/h)\psi^\pm(x,h)$,
where
\begin{equation}
          \label{S-y1y2}
S^\pm(\Lambda^\pm)=\int_0^T P^\pm(t)\,dX^\pm(t).
\end{equation}
Clearly, the correspondence $E^\pm=H|_{\Lambda^\pm}\leftrightarrow
\Lambda^\pm$ is one-to-one. Therefore, $S^\pm$ can be considered
as a function of $E^\pm$, $S^\pm=S^\pm(E^\pm)$. As this dependence
is continuous and monotonic, it can be inverted: $E^\pm=E^\pm(S^\pm)$;
$E^+$ is an increasing function, while $E^-$ is a decreasing one.

Each $\Lambda^\pm$ implies a periodic family of trajectories
$\Lambda^\pm_j:=\Lambda^\pm+(2\pi j,0)$
and corresponding quasimodes
$(E^\pm,\psi^\pm_j)$ with $\psi^\pm_j(x,h)=\exp(2\pi i j x/h)\psi^\pm(x,h)\equiv
T_1^{-j}\psi^\pm(x,h)$.
To construct a Bloch quasimode we use an ansatz similar to that
we use in the previous subsection,
\[
\Psi^\pm(x,\mathbf{k},h)=\sum_{j\in\mathbb{Z}} C^\pm_j \psi^\pm_j(x,h)=
\sum_{j\in\mathbb{Z}} C^\pm_j T_1^{-j}\psi^\pm(x,h).
\]
Therefore,
$C^\pm_{j+1}=C^\pm e^{2\pi i j k_1}$ ($C^\pm$ is a normalizing constant), and
$S^\pm(E^\pm)=h(n\mp k_2),\quad n\in\mathbb{Z}$.
Denote $E^\pm_n(k_2,h):=E^\pm\big(S^\pm=h(n \mp k_2)\big)$;
these functions can be viewed as semiclassical dispersion relations.
Clearly, the functions $E^+_n$
are decreasing functions of $k_2$, while $E^-_\pm$ are increasing ones,
therefore, in some critical points $k_2^*$ one has
$E^*:=E^-_{n}(k_2^*,h)=E^+_m(k_2^*,h)$, as illustrated in figure~\ref{fig-quant}b.
The corresponding points $E^*$
are usually treated as approximations of gaps, more precisely,
one expects that in $o(h^\infty)$-neighborhood of each such value
there is a gap, whose width is also $o(h^\infty)$~\cite{faure}.
The asymptotics of the true dispersion relations
can be combined from pieces of $E^+_n$ and $E^-_m$ (figure~\ref{fig-quant}b).

\bigskip

As we see, we have $o(h^\infty)$-narrow bands in the first case
and $o(h^\infty)$-narrow gaps in the second case.
Our aim is to describe the transition between two these two extremal cases, which correspond
to the semiclassical asymptotics near separatrices.
To obtain at least the first non-trivial term
in the asymptotics of the width of $O(h^\infty)$-small bands and gaps
one should take into account the interaction between
neighboring cells, which is an extremely different problem~\cite{faure2}.
Such a calculation involves topological characteristics of the Hamiltonian, which results
in a description of the quantized Hall conductance~\cite{tknn}.
It is interesting to emphasize that the resulting topological numbers come
from index-like characteristics of a certain path on the Reeb graph~\cite{faure};
tjis provides an additional path index-like interpretation of the
Chern classes, related to their their description through the cycle
related to edges states on the Riemann surface of the Bloch functions~\cite{hts}.

\section{Singular Bohr-Sommerfeld rules and their modification
for the periodic problem}

In this section, we give a short description of the semiclassical asymptotics
near separatrices; this technique was developed in~\cite{cdv-par}.

Let $H(p,x)$ be an arbitrary classical Hamiltonian (not necessary a Harper-like one)
with non-degenerate critical points and $\Hat H_h$ be the corresponding quantum Hamiltonian.
Let $E$ be a critical value of $H$ in the sense that
the level set $\Lambda:=\{(p,x)\in\mathbb{R}^2: H(p,x)=E\}$
contains a saddle point of $H$; our aim is to study the asymptotics
of the spectrum of $\Hat{H}_h$ in the interval $[E-Ah,E+Ah]$, $A>0$.
We assume that $\Lambda$ is a compact connected set.

\subsection{Preliminary constructions}
Near the critical energy it is more convenient to use the scaled energy, i.~e. we are going
to solve the equation
\begin{equation}
            \label{eq-h-psi}
(\Hat H_h-E-\lambda h)\Psi=o(h^\infty),
\end{equation}
where $\lambda$ is a new spectral value to be found;
we consider the situation when $\lambda$ runs through some
finite interval $[-A,A]$, $A>0$. The conditions which guarantee
the existence of such solutions are called the singular
Bohr-Sommerfeld rules.
\begin{figure}\centering
\includegraphics[width=30mm]{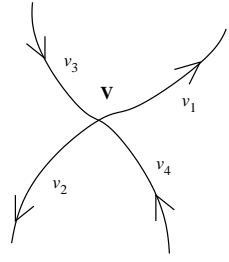}
\caption{Enumeration of edges near vertices}
         \label{fig2-1}
\end{figure}
Due to the non-degeneracy
of the saddle points, $\Lambda$ is a tetravalent graph embedded into the plane
$\mathbb{R}^2_{px}$, and the saddle points are point of branching.
The edges of the graph are smooth curves; each of these curves, $v$, delivers
a part of solution by means of the usual WKB-asymptotics~\cite{mf}; we denote this function by $\psi_v$.
Our aim is to glue these contributions together near saddle points (vertices
of the separatrix graph) in order to obtain a requested solution.
Let us introduce a enumeration of the edges of the separatrix
near each vertex in the following manner: the direct cyclic order
is $v_1,v_4,v_2,v_3$, and the quadrant formed by the edges $v_1$ and $v_3$
is pointed to the top, see figure~\ref{fig2-1}. Near each vertex $\mathbf{V}$
we try to construct the requested solution in the form
$\Psi=x_1^{\mathbf{V}}\psi_{v_1}\oplus x_2^{\mathbf{V}}\psi_{v_2}\oplus
x_3^{\mathbf{V}}\psi_{v_3}\oplus x_3^{\mathbf{V}}\psi_{v_3}$, where
$x_j^{\mathbf{V}}\in\mathbb{C}$, $j=1,2,3,4$. (Sometimes we omit the superscript $\mathbf{V}$
and write simply $x_j$.)
Therefore, an edge can have different numbers near different vertices.
To any vertex $\mathbf{V}$ of the graph we assign a so-called \emph{semiclassical
invariant} $\varepsilon^{\mathbf{V}}$ which is a formal power series in $h$,
\[
\varepsilon^{\mathbf{V}}(\lambda,h)=\sum_{j=0}^\infty \varepsilon^{\mathbf{V}}_j(\lambda) h^j,
\quad\text{where}\quad
\varepsilon^{\mathbf{V}}_0(\lambda)=\pm\frac{\lambda}{\sqrt{|\det H''(\mathbf{V})|}},
\quad H''=\begin{pmatrix}
H_{pp} & H_{px}\\ H_{xp} & H_{xx}
\end{pmatrix}
\]
and the sign $\pm$ coincides with the sign of $H$
in the quadrants formed by the edges 1 and 3.

Each cycle $\gamma$ on the graph $\Lambda$
will be accompanied by the following three characteristics (see~\cite{cdv-par} for details):
\begin{itemize}
\item Principal action $A_\gamma$,
\[A_{\gamma}=\oint_\gamma p\ dx,\]
\item Renormalized time $I_\gamma$. For cycles $\gamma$ crossing
critical points
with corners, we put
\[
I_\gamma =\vp\oint_\gamma dt:=
\sum_{j=1}^n \int_{\mathbf{c}_j}^{\mathbf{c}_{j+1}} dt,
\quad \mathbf{c}_j\in\gamma,\quad \mathbf{c}_1=\mathbf{c}_{n+1},
\]
where the points $\mathbf{c}_j$ are chosen in such a way that
each of the pieces $(\mathbf{c}_j,\mathbf{c}_{j+1})\subset\gamma$ contains exactly
one corner $\mathbf{V}_j$, and the integrals are calculated as
\[
\int_{\mathbf{c}_j}^{\mathbf{c}_{j+1}} dt:=\lim_{\mathbf{a},\mathbf{b}\to \mathbf{V}_j}
\bigg(
\int_{\mathbf{c}_j}^\mathbf{a} dt +\int_\mathbf{b}^{\mathbf{c}_{j+1}} dt +
s_j\log\Big|
\int_{R_{\mathbf{ab}}} dp\wedge dx\Big|\bigg),
\]
where $s_j:=\mp\lambda/\sqrt{|\det H''(\mathbf{V}_j)|}$,
and $R_{\mathbf{ab}}$ is a parallelogram spanned by the points
$\mathbf{a}$, $\mathbf{V}_j$, and $\mathbf{b}$,
and the sign $\mp$ is "--" if the direction of integration corresponds
to $dt$ and "+" otherwise.
By $dt$ we denote the Hamiltonian time, $dt(\sgrad H)=1$.
This definition of $I_\gamma$ is then extended by additivity
to all cycles (not necessary with corners).
\item Maslov index $m_\gamma$. If none of the tangent 
to $\gamma$ at the corners is parallel to the $p$-axis,
the index $m_\gamma$ is calculated as the sum of the indices
of all the turning points in which $\gamma$ is smooth.
If there are tangents parallel to the $p$-axis, one can destroy
them by small variation; this must be reflected in the enumeration
of edges near all the vertices.
\end{itemize}

\subsection{Singular Bohr-Sommerfeld rules}
The singular Bohr-Sommerfeld rules, which finally will result
in conditions for $\lambda$, come from the following procedure.

Near each saddle point $\mathbf{V}$
there exist a canonical transformation
$\chi(q,y)=(p,x)$, $\chi(\mathbf{O})=\mathbf{V}$, such that $H(p,x)=W(qy)qy$
with a certain smooth function $W$.
This implies an elliptic Fourier integral operator $\Hat U$
and a pseudodifferential operator $\Hat W$, elliptic at the origin,
such that
$\Hat H\Hat U=\Hat U \Hat W(\widehat{yq}-h\varepsilon^{\mathbf{V}})$.
Using this representation one can construct a microlocal basis $\psi_j$,
$j=1,2,3,4$,
of semiclassical solutions near $\mathbf{V}$. More precisely,
we put
\begin{equation}
              \label{eq-model}
\begin{aligned}
\varphi_1(y)&=B^\mathbf{V}\frac{Y(y)}{\sqrt{|y|}}\ e^{i\varepsilon\log|y|},
&
\varphi_2(y)&=B^\mathbf{V}\frac{Y(-y)}{\sqrt{|y|}}\ e^{i\varepsilon\log|y|},\\
\varphi_3(y)&=B^\mathbf{V}
\frac{e^{-i\pi/4}}{\sqrt{2\pi h}}
\int_{\mathbb{R}}\frac{Y(t)}{\sqrt{|t|}}\ e^{iyt/h}\ e^{-\varepsilon\log|t|}dt,
&
\varphi_4(y)&=B^\mathbf{V}\frac{e^{-i\pi/4}}{\sqrt{2\pi h}}
\int_{\mathbb{R}}\frac{Y(-t)}{\sqrt{|t|}}\ e^{iyt/h}\ e^{-\varepsilon\log|t|}dt,
\end{aligned}
\end{equation}
where $Y$ is the Heaviside function,
and then $\psi_j=\Hat U\varphi_j$ is a basis element corresponding
to the edge $j$, $j=1,2,3,4$. Here $B^\mathbf{V}$ is a normalizing constant.

A linear combination $x_1\psi_1\oplus x_2\psi_2\oplus x_3\psi_3\oplus
x_4\psi_4$, $x_j\in\mathbb{C}$, $j=1,2,3,4$, which extends the WKB-solution
to the saddle point, defines a function near $\mathbf{V}$ iff $x_j$'s
satisfy the linear system
\begin{equation}
           \label{eq-match}
\begin{pmatrix}x_3\\x_4\end{pmatrix}=
\mathcal{E}\begin{pmatrix}
1 & ie^{-\varepsilon\pi}\\
ie^{-\varepsilon\pi} &1
\end{pmatrix}\begin{pmatrix}x_1\\x_2\end{pmatrix},
\qquad
\mathcal{E}:=
\frac{1}{\sqrt{1+e^{-2\pi\varepsilon}}}e^{i\arg\Gamma(\frac{1}{2}+i\varepsilon)+
i\varepsilon\log h},
\qquad \varepsilon=\varepsilon^{\mathbf{V}}.
\end{equation}
Clearly, $x_j$'s corresponding to different vertices must be connected
with each other in a certain sense.
To describe this correspondence we cut several edges in order to
get a maximal tree on the separatrix. Now consider an arbitrary edge
$v$ between two vertices $\mathbf{V}'$ and $\mathbf{V}''$.
Denote the corresponding
$x$-coefficients by $x'_j$ and $x''_j$  respectively, $j=1,2,3,4$,
and assume that $v$ has index $j$ with respect to $\mathbf{V}'$ and index
$k$ with respect to $\mathbf{V}''$. 
We put
\begin{equation}
            \label{eq-hol}
x'_j=x''_k, \quad\text{ if $v$ is not cut,}\qquad
x'_j=\hol \gamma\  x''_k, \quad\text{ if $v$ is cut},
\end{equation}
where $\hol\gamma$ is the so-called holonomy of the cycle $\gamma$
formed by the edge $e$ and the edges of the maximal tree (this cycle
is unique). For the holonomy holds the estimate $|\hol\gamma|=1$,
$\Arg\hol\gamma=A_\gamma/h+\lambda I_\gamma+\pi m_\gamma/2 +O(h)$.
Note that the first equality in~\eqref{eq-hol} fix
the constants $B^{\mathbf{V}}$ uniquely (up to a common multiplicator).

The equations~\eqref{eq-match} and~\eqref{eq-hol} written
for all the vertices and all the cycles respectively compose a linear system
for the coefficients $x^\mathbf{V}_j$, $j=1,2,3,4$.
The condition for the existence of non-trivial solutions
(non-vanishing determinant) is called the singular Bohr-Sommerfeld rules.

\subsection{Periodic problem}

Our aim is to apply the singular Bohr-Sommerfeld rules
to the Harper-like  operators. The problem is that the level set
of $E$ is always periodic and, respectively, unbounded.
To apply this technique to the problem in question
we  take into account the Bloch conditions already at the stage
of the construction of the solution (in contrast to the
smooth case), in other words,
we are going to apply the procedure described above for constructing
the quasimode $\big(\Psi(x,\mathbf{k},h),E(\mathbf{k},h)\big)$ such that
$\Psi$ satisfies~\eqref{eq-bloch} and~\eqref{eq-h-psi}.
Clearly, this condition can be rewritten as the set of equalities
$B^{\mathbf{V}+2\pi\mathbf{l}}=B^\mathbf{V}$,
and $x_j^{\mathbf{V}+2\pi\mathbf{l}}= e^{i\langle\mathbf{l}|\mathbf{k}\rangle}x^\mathbf{V}_j$,
$j=1,2,3,4$, for all vertices $\mathbf{V}$
and $\mathbf{l}\in\mathbb{Z}^2$, where $B^\mathbf{V}$ are constants
from~\eqref{eq-model}. This clarifies the meaning of the maximal tree
in the periodic case: Instead of constructing a maximal tree in the whole plane
it is sufficient to find a maximal
tree on the reduced phase space $\mathbb{T}^2_{px}$; all the
cycles will be viewed then as cycles on the torus.

\section{Transition between finite and infinite motions}\label{sec4}

In this section, we consider spectral regions
corresponding to the transition between closed and open trajectories.
In this case the separatrices are non-compact in one direction only,
and we assume that they are directed along the $x$-axis.
Clearly, one can proceed first exactly in the same way
as in the case of open trajectories, which will result in the following
ansatz for the requested solution:
\[
\Psi(x,\mathbf{k},h)=c\sum_{j\in\mathbb{Z}}e^{ijk_1}T_1^{-j}\psi(x,k_2,h),
\]
where $\psi$ must be a quasimode associated with the separatrix,
$(\Hat H_h-E-\lambda h)\psi=o(h^\infty)$,
and satisfying $T_2\psi=e^{i k_2}\psi$. It is clear that
the corresponding
semiclassical dispersion relations will not depend on $k_1$.

\subsection{Two edges of infinite motion and one edge of finite motion}\label{subsec41}
\begin{figure}
\begin{minipage}{100mm}\centering
\includegraphics[height=60mm]{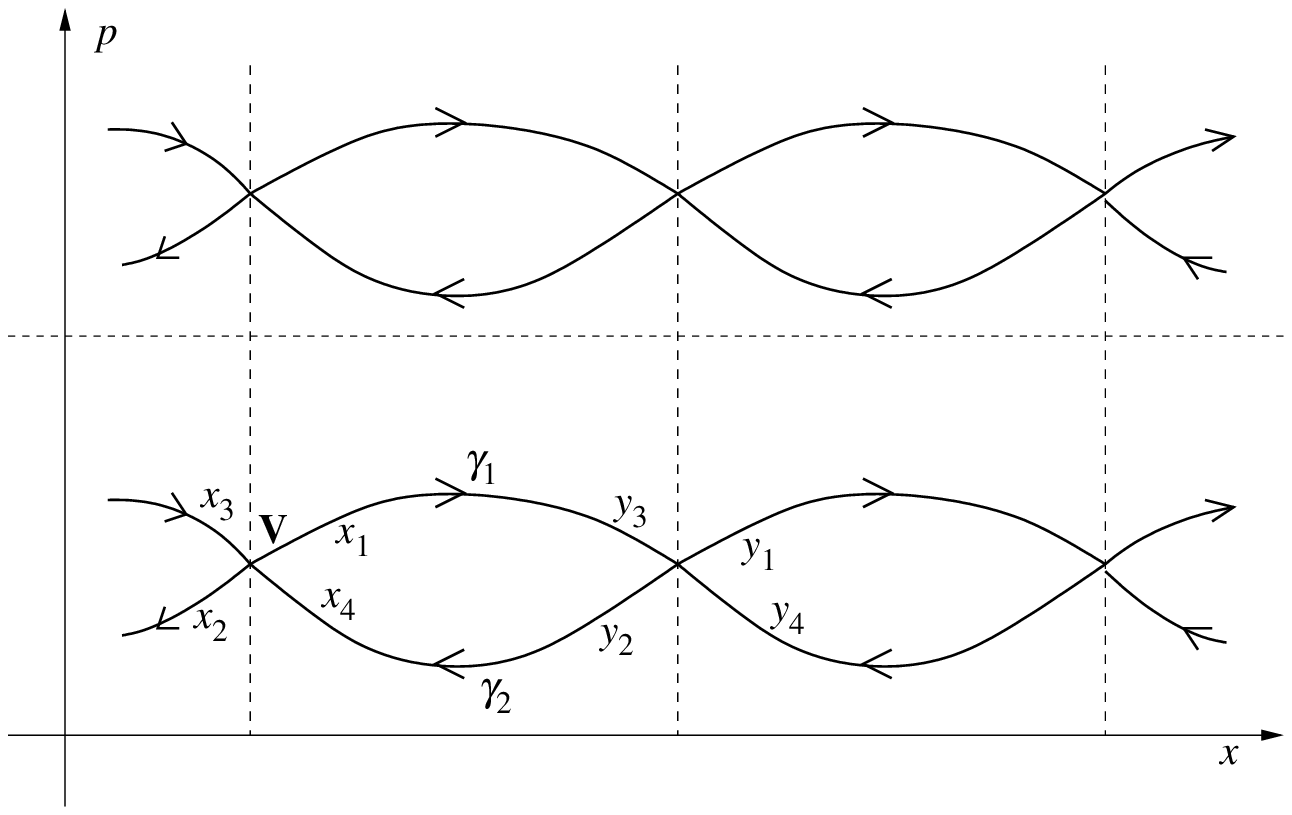}\\
(a)
\end{minipage}
\begin{minipage}{40mm}\centering
\includegraphics[height=60mm]{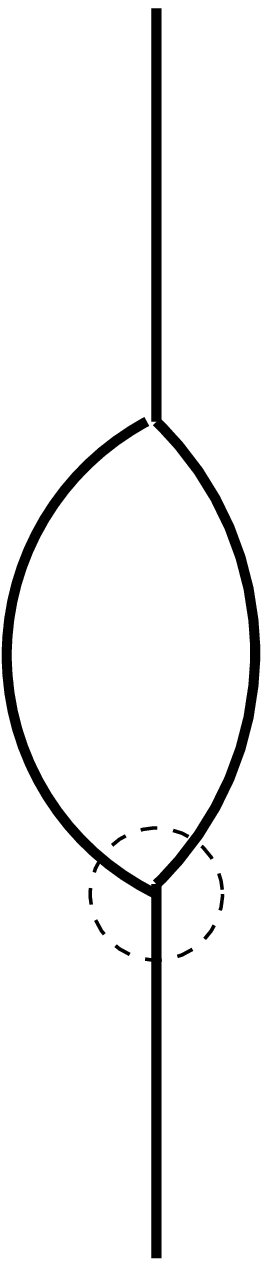}\\
(b)
\end{minipage}
\caption{Transition between one edge of finite motion and two edges of infinite motion:
 (a) Structure of the separatrix, (b)
The corresponding part of the Reeb graph}
\label{fig1}
\end{figure}

In this subsection, we consider probably the simplest structure
of the separatrix. More precisely, we assume that the energy level
$H=E$ on the reduced phase space $\mathbb{T}^2_{px}$ contains exactly one
critical point, to be denoted by $\mathbf{V}$. The corresponding
separatrix on the plane $\mathbb{R}^2_{px}$ has then the shape showed
in figure~\ref{fig1}. In terms of the Reeb graph this situation means
that we consider a transition between two edges of infinite motions
and one edge of finite motion. Near the corresponding branching point
the Reeb graph has a $Y$-like shape. (We assume that the closed trajectories
lie under the critical energy level.)
Denote the semiclassical invariant of $\mathbf{V}$
by $\varepsilon(\lambda,h)$. Clearly,
\begin{equation}
       \label{eps-1}
\varepsilon(\lambda,h)=\lambda/w+O(h),\qquad
w:=\sqrt{|\det H''(\mathbf{V})|}.
\end{equation}
The separatrix on $\mathbb{T}^2_{px}$ is a graph with one vertex
and two edges, $\gamma_1$ and $\gamma_2$, which are cycles on the torus,
see figure~\ref{fig1}.
In order to obtain a maximal tree one has to cut both edges.
Denote $\hol\gamma_j=:e^{i\alpha_j}$, $j=1,2$, where
\[
\alpha_j=\frac{A_j}{h}+\lambda I_j+\frac{\pi m_j}{2}+O(h),\qquad
A_j:=\oint_{\gamma_j}p\ dx,
\quad
I_j=\vp \int_{\gamma_j} dt,
\quad m_j=\Ind\gamma_j,
\quad j=1,2.
\]
To simplify the notation we put $x_j:=x^{\mathbf{V}}_j$,
$y_j:=x^{\mathbf{V}+(0,2\pi)}_j$, $j=1,2,3,4$.
The singular Bohr-Sommerfeld rules and the Bloch conditions
lead us to the following equations:
\begin{gather*}
\text{Bloch-periodicity conditions: }\quad
y_j =e^{i k_2} x_j, \quad j=1,2,3,4,\\
\text{Matching at $\mathbf{V}$: }
\left\{\begin{aligned}
x_3&=\mathcal{E}x_1 +i e^{-\varepsilon\pi}\mathcal{E}x_2\\
x_4&=i e^{-\varepsilon\pi}\mathcal{E} x_1 +\mathcal{E}x_2,
\end{aligned}\right.\qquad
\text{Holonomy equations: }
\left\{
\begin{aligned}
y_3&=e^{i\alpha_1} x_1, & \text{ for } \gamma_1,\\
x_4&=e^{i\alpha_2} y_2, & \text{ for } \gamma_2.
\end{aligned}\right.
\end{gather*}
where
\[
\mathcal{E}:=
\frac{1}{\sqrt{1+e^{-2\pi\varepsilon}}}e^{i\arg\Gamma(\frac{1}{2}+i\varepsilon)+
i\varepsilon\log h},\\
\]
The condition of existence of non-zero solutions is equivalent
then to the equation
\[
\cos\Big(
\arg\Gamma\big(\frac{1}{2}+i\varepsilon\big)+\varepsilon\log h
-\frac{\alpha_1+\alpha_2}{2}\Big)=
\frac{1}{\sqrt{1+e^{-2\pi\varepsilon}}}
\cos\Big(k_2-\frac{\alpha_2-\alpha_1}{2}\Big).
\]
The spectral parameter $\lambda$ enters this equation through $\varepsilon$,
$\alpha_{1}$ and $\alpha_2$. 
One has obviously
\[
\arg\Gamma\big(\frac{1}{2}+i\varepsilon\big)+\varepsilon\log h
-\frac{\alpha_1+\alpha_2}{2}
=\pm\arccos
\frac{1}{\sqrt{1+e^{-2\pi\varepsilon}}}\,
\cos\Big(k_2-\frac{\alpha_2-\alpha_1}{2}\Big)
+2\pi n,\quad
n\in\mathbb{Z}.
\]

\noindent To find approximate solutions we represent $\lambda$ as
$\lambda=\lambda_0+\mu$ with $\mu=o(1)$; in this way we can solve
the equation near each $\lambda_0$.
Taking into account Eq.~\eqref{eps-1} one can write
\begin{multline*}
\arg\Gamma\Big(\frac{1}{2}+\frac{i\lambda_0}{w}\Big)+
\frac{\lambda_0}{w}\log h+\frac{\mu}{w}\log h-\frac{(A_1+A_2)}{2h}
-\frac{I_1+I_2}{2}\lambda_0-\frac{\pi(m_1+m_2)}{2}\\
{}=\pm\arccos\frac{1}{\sqrt{1+e^{-2\pi\lambda_0/w}}}
\cos\Big(k_2-\frac{A_2-A_1}{2h}-\frac{I_2-I_1}{2}\lambda_0\Big)+2\pi n
+O(\lambda)+O(h\log h),\quad n\in\mathbb{Z}
\end{multline*}
(one can show easily that $m_1-m_2=0$),
or, finally,
\begin{equation}
             \label{eq-disp1}
\begin{gathered}
\mu=\mu^\pm_n(k_2,\lambda_0,h)=
\frac{w}{\log h}\,\bigg\{
N(\lambda_0,h)
\pm\arccos\dfrac{\cos\big(k_2-\Delta(\lambda_0,h)\big)}{\sqrt{1+e^{-2\pi\lambda_0/w}}}
+2\pi n\bigg\}\,\bigg[1+O\Big(\frac{1}{\log h}\Big)\bigg]+O(h)\\
{}=\frac{w}{\log h}\times\bigg\{
N(\lambda_0,h)
\pm\arccos\dfrac{\cos\Big(k_2-\Delta(\lambda_0,h)\Big)}{\sqrt{1+e^{-2\pi\lambda_0/w}}}
+2\pi n\bigg\}+o\Big(\frac{1}{\log h}\Big),
\end{gathered}
\end{equation}
where
\begin{gather*}
N(\lambda_0,h)=\frac{A_1+A_2}{2h}+\frac{\lambda_0(I_1+I_2)+\pi(m_1+m_2)}{2}-\frac{\lambda_0\log h}{w}
-\arg\Gamma\Big(\frac{1}{2}+\frac{i\lambda_0}{w}\Big),\\
\Delta(\lambda_0,h)=\dfrac{A_2-A_1}{2h}+\lambda_0\dfrac{I_2-I_1}{2}
\end{gather*}
and $n$ are integers such that the expression in the curly brackets in~\eqref{eq-disp1} is 
$o(\log h)$.
Returning to the original spectral parameter we obtain a series
of semiclassical dispersion relations: in a $o(h)$-neighborhood
of $E+\lambda_0h$ they take the form
$E^\pm_n(k_1,k_2,\lambda_0,h)=E+\lambda_0 h +h \mu^\pm_n(k_2,\lambda_0,h)$.
The expression obtained can be used for estimating the band- and the gapwidth.
More precisely, the bands and the gaps in a $o(h)$-neighborhood of $E+\lambda_0h$
have the width
\begin{gather}
          \label{eq-b1}
B(\lambda_0,h)=\frac{2wh}{|\log
h|}\arcsin\frac{1}{\sqrt{1+e^{-2\pi\lambda_0/w}}}
+o\Big(\frac{h}{\log h}\Big),\\
          \label{eq-g1}
G(\lambda_0,h)=\frac{2wh}{|\log h|}\arccos\frac{1}{\sqrt{1+e^{-2\pi\lambda_0/w}}}
+o\Big(\frac{h}{\log h}\Big)
\end{gather}
respectively. In particular,
for $\lambda_0=0$ the bands and the gaps have approximately the
same width $\pi w h/(2|\log h|)$.

\subsection{Two edges of infinite motion and two edges of finite motion}\label{subsec42}
\begin{figure}
\begin{minipage}{100mm}\centering
\includegraphics[height=60mm]{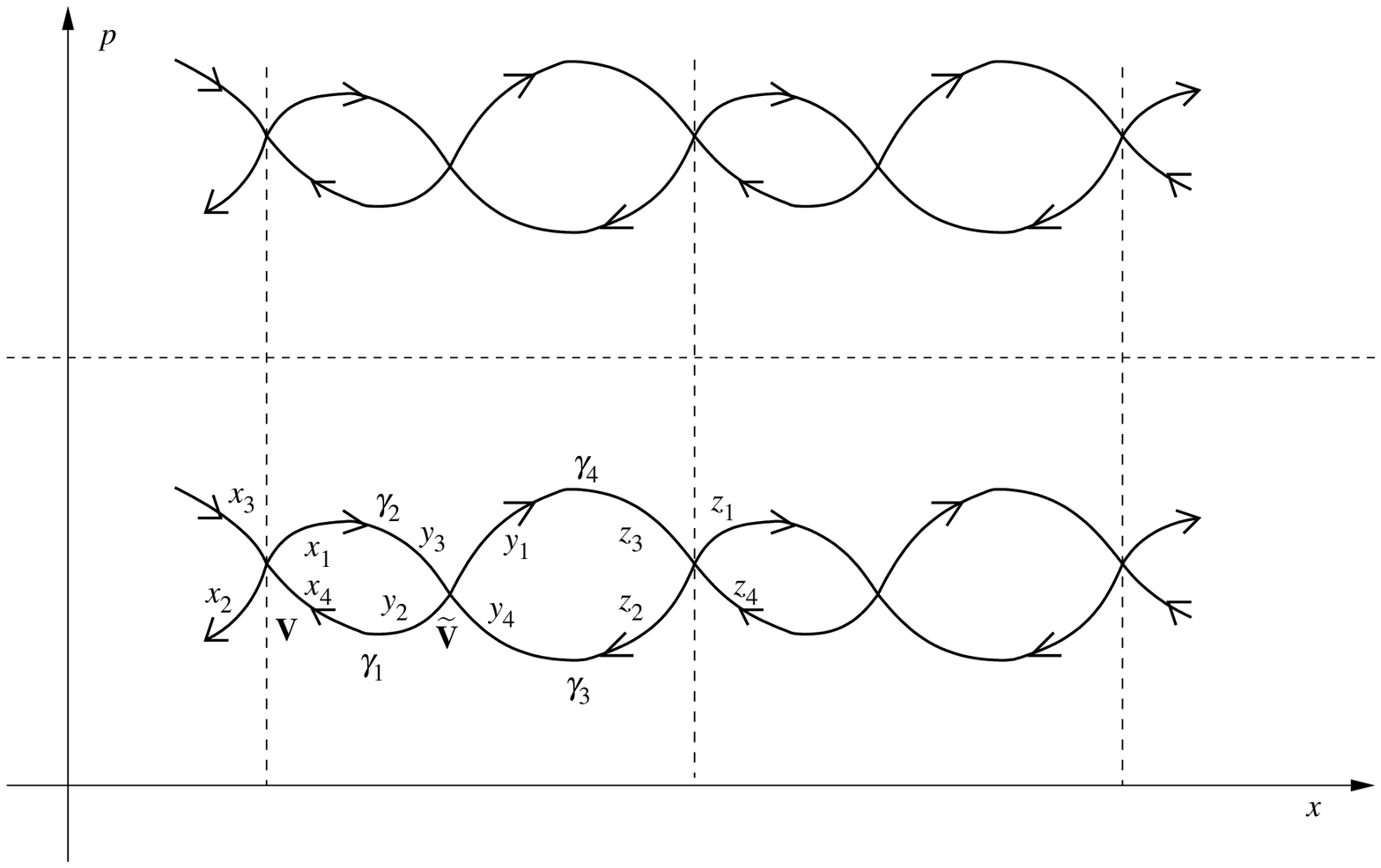}\\(a)
\end{minipage}
\begin{minipage}{40mm}\centering
\includegraphics[height=60mm]{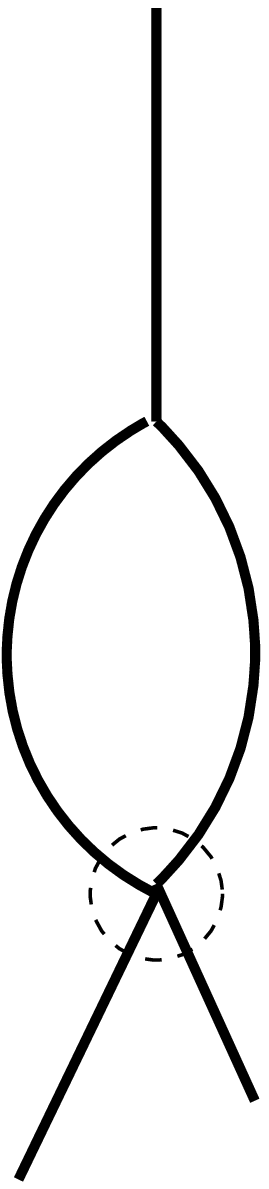}\\(b)
\end{minipage}
\caption{Transition between two edges of finite motion and
two edges of infinite motion: (a)
Structure of the separatrix, (b)
The corresponding part of the Reeb graph.}
         \label{fig2}
\end{figure}

We consider now another structure of the separatrix.
Let us assume that the energy level $H=E$ on the reduced
phase space is a connected set containing two critical
points, which we denote by $\mathbf{V}$ and
$\Tilde{\mathbf{V}}$. The separatrix
has the shape sketched in figure~\ref{fig2}. The Reeb graph has near
the corresponding branching point a $X$-like shape,
where the two upper edges correspond to open trajectories.

Denote the semiclassical invariants of $\mathbf{V}$ and $\Tilde{\mathbf{V}}$
by $\varepsilon$ and $\Tilde\varepsilon$ respectively,
\begin{equation}
       \label{eps-2}
\varepsilon(\lambda,h)=\lambda/w+O(h),
\quad \Tilde\varepsilon=\lambda/\Tilde w+O(h),
\qquad
w:=\sqrt{\big|\det H''(\mathbf{V})\big|},
\quad
\Tilde w:=\sqrt{\big|\det H''(\Tilde{\mathbf{V}})\big|},
\end{equation}
and put
\begin{equation}
                \label{eq-EPS}
\mathcal{E}:=\frac{\exp\left[ i\arg\Gamma\Big(\frac12+i\varepsilon\Big)+i\varepsilon\log h,
\right]}{\sqrt{\displaystyle 1+e^{-2\pi\varepsilon}}},\qquad
\Tilde{\mathcal{E}}:=\frac{\exp\left[
i\arg\Gamma\Big(\frac12+i\Tilde\varepsilon\Big)+i\Tilde\varepsilon\log h,
\right]}{\sqrt{\displaystyle 1+e^{-2\pi\Tilde\varepsilon}}}\end{equation}
In order to obtain a maximal tree on the reduced phase space we cut all
the edges but $\gamma_1$, then one get three cycles: $\Tilde\gamma_1=\gamma_1+\gamma_2$,
$\Tilde\gamma_2=\gamma_4-\gamma_1$, and $\Tilde\gamma_3=\gamma_1+\gamma_3$;
put $\alpha_j:=\arg\hol \Tilde\gamma_j$, $j=1,2,3$.

Denote $x_j:=x^{\mathbf{V}}_j$, $y_j:=x^{\Tilde{\mathbf{V}}}_j$,
$z_j:=x^{\mathbf{V}+(0,2\pi)}_j$, $j=1,2,3,4$, then the quantization
conditions are:
\begin{gather*}
\text{Bloch-periodicity conditions:}
 \quad z_j=e^{i k_2} x_j, \quad j=1,2,3,4,\\
\text{Matching conditions at $\mathbf{V}$: }
\left\{
\begin{aligned}
x_3&=\mathcal{E} x_1+ie^{-\varepsilon \pi}\mathcal{E} x_2,\\
x_4&=ie^{-\varepsilon\pi}\mathcal{E} x_1+\mathcal{E} x_2,
\end{aligned}\right.
\quad\text{at $\Tilde{\mathbf{V}}$: }
\left\{
\begin{aligned}
y_3&=\Tilde{\mathcal{E}}y_1+ie^{-\Tilde\varepsilon\pi}\Tilde{\mathcal{E}}y_2,\\
y_4&=ie^{-\Tilde\varepsilon\pi}\Tilde{\mathcal{E}}y_1+\Tilde{\mathcal{E}}y_2,
\end{aligned}
\right.\\
\text{Holonomy equations: }
y_2=x_4,\quad
y_3=e^{i\alpha_1} x_1,\quad
z_3=e^{i\alpha_2} y_1,\quad
y_4=e^{i\alpha_3} z_2.
\end{gather*}

The condition of the existence of non-trivial solutions leads to
the equation
\begin{multline}
                \label{eq-alpha}
\cos\bigg(
\arg\Gamma\Big(\frac{1}{2}+i\varepsilon\Big)+
\arg\Gamma\Big(\frac{1}{2}+i\Tilde{\varepsilon}\Big)+
(\varepsilon+\Tilde\varepsilon)\log h-\frac{\alpha_1+\alpha_2+\alpha_3}{2}
\bigg)\\
{}=
\dfrac{1}{\displaystyle\sqrt{(1+e^{-\varepsilon\pi})(1+e^{-\Tilde\varepsilon\pi})}}
\bigg(
\cos\Big(k_2-\frac{\alpha_1+\alpha_2-\alpha_3}{2}\Big)
-e^{-(\varepsilon+\Tilde\varepsilon)\pi}\cos\frac{\alpha_1-\alpha_2-\alpha_3}{2}
\bigg).
\end{multline}
The parameter $\lambda$ enters this equation through $\varepsilon$, $\Tilde\varepsilon$,
and $\alpha_j$, $j=1,2,3$.

To obtain a more clear picture, let us introduce
holonomies of the edges as solutions of the following equalities:
$\alpha_1=\beta_1+\beta_2$, $\alpha_2=\beta_4-\beta_1$,
$\alpha_3=\beta_1+\beta_3$, $\arg\hol(\Tilde\gamma_2+\Tilde\gamma_4)=\beta_2+\beta_4$;
these solutions can be represented as
\[
\beta_j=\frac{B_j}{h}+\lambda J_j+\frac{\pi m_j}{2}+O(h),
\quad
B_j=\int_{\gamma_j}p\ dx,
\quad m_j=\Ind\gamma_j,
\quad j=1,2,3,4,
\]
then \eqref{eq-alpha} takes the form
\begin{multline}
               \label{eq-beta}
\arg\Gamma\Big(\frac{1}{2}+i\varepsilon\Big)+
\arg\Gamma\Big(\frac{1}{2}+i\Tilde{\varepsilon}\Big)+
(\varepsilon+\Tilde\varepsilon)\log h
-\frac{\beta_1+\beta_2+\beta_3+\beta_4}{2}\\
{}\quad=
\pm\arccos
\dfrac{\cos\Big(k_2-\frac{(\beta_2+\beta_4)-(\beta_1+\beta_3)}{2}\Big)
-e^{-(\varepsilon+\Tilde\varepsilon)\pi}\cos\frac{(\beta_1+\beta_2)-(\beta_3+\beta_4)}{2}}
{\displaystyle\sqrt{(1+e^{-\varepsilon\pi})(1+e^{-\Tilde\varepsilon\pi})}}
+2\pi n,\quad
n\in\mathbb{Z}.
\end{multline}
The quantities $\beta_j$ can be viewed as ``weights'' of the separatrice edges,
then Eq.~\eqref{eq-beta} shows how the relationship between them influences
the dispersion relations.

Like in the previous case we represent $\lambda$ as $\lambda_0+\mu$, $\mu=o(1)$,
and solve (approximately) the equation for $\mu$. The solutions take the form
\begin{equation}
              \label{eq-disp2}
\begin{gathered}
\mu=\mu_n^\pm(k_2,\lambda_0,h)=
\frac{w\Tilde{w}}{(w+\Tilde w)\log h}\times\bigg\{
N(\lambda_0,h)\\
{}\pm\arccos\frac{\cos\big(k_2-\Delta_2(\lambda_0,h)\big)-e^{-(1/w+1/\Tilde w)\lambda_0\pi}
\cos\Delta_1(\lambda_0,h)}{\sqrt{(1+e^{-2\pi\lambda_0/w})(1+e^{-2\pi\lambda_0/\Tilde w})}}
+2\pi n\bigg\}\times\bigg(1+O\Big(\frac{1}{\log h}\Big)\bigg)+O(h),
\end{gathered}
\end{equation}
where
\begin{align*}
N(\lambda_0,h)&:=
\frac{B_1+B_2+B_3+B_4}{2h}+\lambda_0\frac{J_1+J_2+J_3+J_4}{2}+\frac{\pi(m_1+m_2+m_3+m_4)}{2}\\
&\qquad{}-\arg\Gamma\Big(\frac{1}{2}+\frac{i\lambda_0}{w}\Big)
-\arg\Gamma\Big(\frac{1}{2}+\frac{i\lambda_0}{\Tilde
w}\Big)-\lambda_0\Big(\frac{1}{w}+\frac{1}{\Tilde w}\big)\log h,\\
\Delta_1(\lambda_0,h)&:=\dfrac{(B_1+B_2)-(B_3+B_4)}{2h}+\lambda_0\dfrac{(J_1+J_2)-(J_3+J_4)}{2},\\
\Delta_2(\lambda_0,h)&:=
\frac{(B_2+B_4)-(B_1+B_3)}{2h}+\lambda_0\frac{(J_2+J_4)-(J_1+J_3)}{2}
\end{align*}
and $n\in\mathbb{Z}$, $n=[N(\lambda_0,h)]+o(\log h)$.
The dispersion relations are
$E(k_1,k_2,h)=E+h\lambda_0+h\mu^\pm_n(k_2,\lambda_0,h)$.
In contrast to the previous situation,
band and gaps have, generally speaking, different
width. More precisely, near the point $E+\lambda_0 h$ one has
bands having the width
\begin{multline}
               \label{eq-b2}
B(\lambda_0,h)=\frac{w\Tilde w h}{(w+\Tilde w)|\log h|}
\bigg[\arcsin\frac{1+e^{-(1/w+1/\Tilde w)\lambda_0\pi}\cos\ \Delta_1(\lambda_0,h)}{
\sqrt{(1+e^{-2\pi\lambda_0/w})(1+e^{-2\pi\lambda_0/\Tilde w})}}\\
{}+\arcsin\frac{1-e^{-(1/w+1/\Tilde w)\lambda_0\pi}\cos\ \Delta_1(\lambda_0,h)}{
\sqrt{(1+e^{-2\pi\lambda_0/w})(1+e^{-2\pi\lambda_0/\Tilde w})}}
\bigg]+o\Big(\frac{h}{\log h}\Big),
\end{multline}
and two groups of gaps having the width
\begin{gather}
               \label{eq-g2a}
G_1(\lambda_0,h)=\frac{2w\Tilde w h}{(w+\Tilde w)|\log
h|}\arccos\frac{1+e^{-(1/w+1/\Tilde w)\lambda_0\pi}\cos\ \Delta_1(\lambda_0,h)}{
\sqrt{(1+e^{-2\pi\lambda_0/w})(1+e^{-2\pi\lambda_0/\Tilde w})}} +o\Big(\frac{h}{\log h}\Big),\\
               \label{eq-g2b}
G_2(\lambda_0,h)=\frac{2w\Tilde w h}{(w+\Tilde w)|\log
h|}\arccos\frac{1-e^{-(1/w+1/\Tilde w)\lambda_0\pi}\cos\ \Delta_1(\lambda_0,h)}{
\sqrt{(1+e^{-2\pi\lambda_0/w})(1+e^{-2\pi\lambda_0/\Tilde w})}}
+o\Big(\frac{h}{\log h}\Big),
\end{gather}
and they come in groups $\dots,B,G_1,B,G_2,\dots$,
The formulas clearly show the fast decaying of the bandwidth for negative $\lambda_0$
and of the gapwidth for positive $\lambda_0$.
From the other side, the ratio bandwidth/gapwidth depends
crucially on the relationship between $B_j$, $J_j$, $\lambda_0$, and $h$.

\section{Transition between topologically different
finite motions (degenerate case)}\label{sec5}

\begin{figure}
\begin{minipage}{100mm}\centering
\includegraphics[height=60mm]{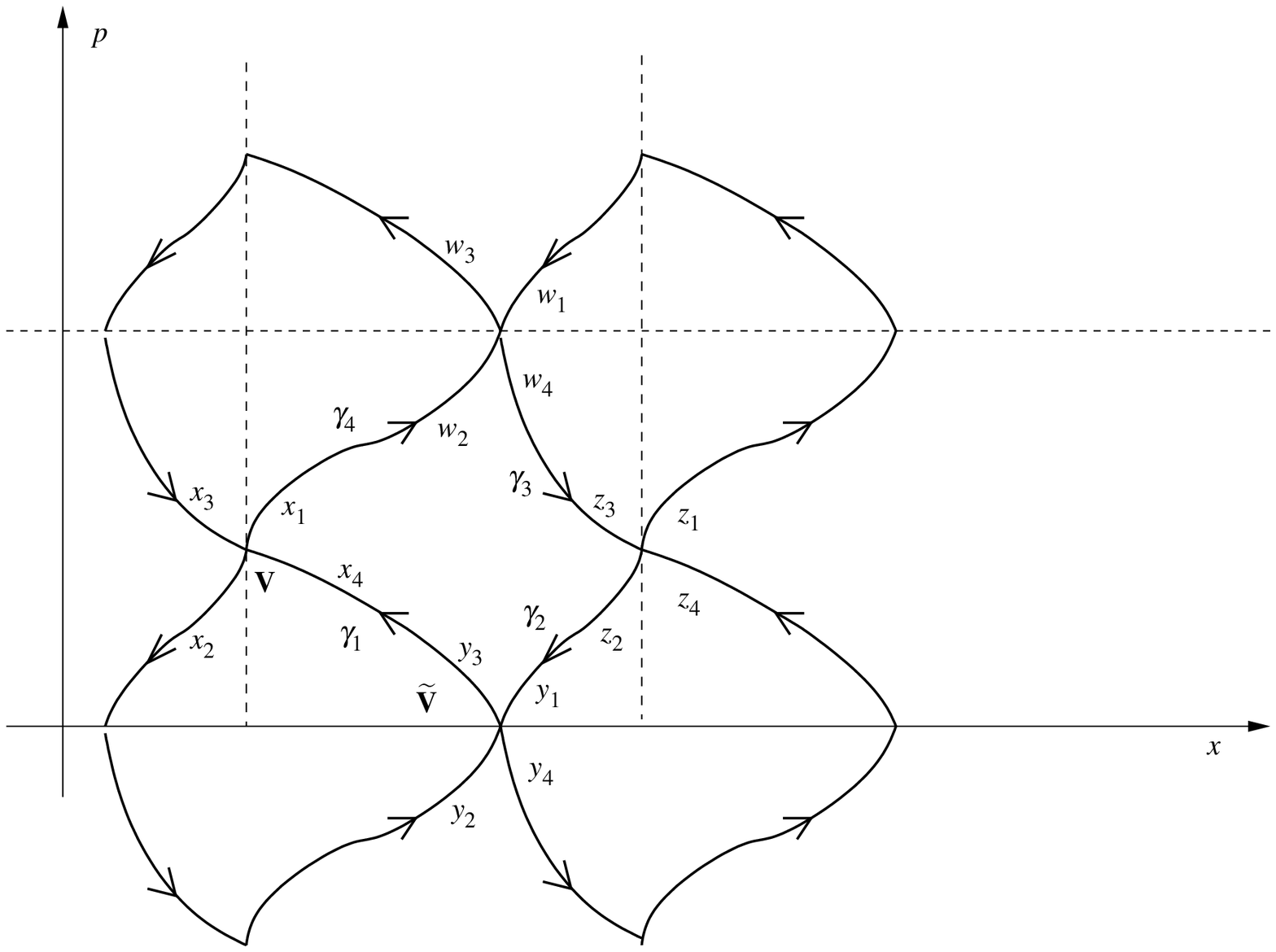}\\(a)
\end{minipage}
\begin{minipage}{40mm}\centering
\includegraphics[height=50mm]{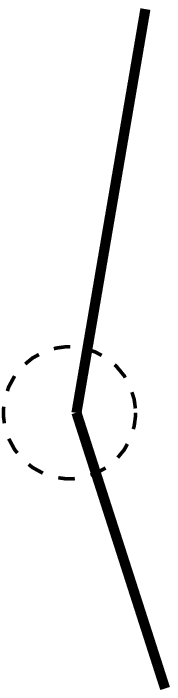}\rule{0mm}{60mm}\\(b)
\end{minipage}
\caption{Transition between two edges of finite motion: (a)
Structure of the separatrix, (b)
The corresponding part of the Reeb graph.}
\label{fig3}
\end{figure}

In this section we consider a situation when all the smooth
trajectories of the classical Hamiltonian are closed.
This situation takes place if, for example, $H$ is invariant
under linear transformation having no real eigenvectors
(rotation by $\pi/2$, for example).
More precisely, we assume that the level set $H=E$ on the reduced
space contains two critical points, which we denote by $\mathbf{V}$
and $\Tilde{\mathbf{V}}$, and the corresponding separatrix in the plane
is non-compact in all directions. This situation is illustrated
in figure~\ref{fig3}. We can expect that the semiclassical dispersion relations in this
case depend on both quasimomenta $k_1$ and $k_2$.
Denote by $\varepsilon$ and $\Tilde\varepsilon$ the semiclassical invariants
of $\mathbf{V}$ and $\Tilde{\mathbf{V}}$ respectively, and use the notation
of~\eqref{eq-EPS}.
Clearly,
\[
\varepsilon=\lambda/w+O(h),\quad \Tilde\varepsilon=-\lambda/\Tilde w+O(h),
\qquad w=\sqrt{\big|\det H''(\mathbf{V})\big|},\quad
\Tilde w=\sqrt{\big|\det H''(\Tilde{\mathbf{V}})\big|}.
\]
An essential point is that the main terms of $\varepsilon$
and $\Tilde\varepsilon$ have opposite signs.

We fix a maximal tree by cutting all the edges but $\gamma_1$, then
three cycles appear: $\Tilde\gamma_1=\gamma_4+\gamma_1$,
$\Tilde\gamma_2=\gamma_3-\gamma_1$, and $\Tilde\gamma_2=\gamma_2+\gamma_1$
with holonomies $e^{i\alpha_j}$, $j=1,2,3$, respectively.
Put $x_j:=x^{\mathbf{V}}_j$, $y_j:=x^{\Tilde{\mathbf{V}}}_j$,
$z_j:=x^{\mathbf{V}+(0,2\pi)}_j$, $w_j:=x^{\Tilde{\mathbf{V}}+(2\pi,0)}_j$,
$j=1,2,3,4$, then we come to the following set of equalities:
\begin{gather*}
\text{Bloch-periodicity conditions: }
w_j=e^{i k_1}y_j,\quad z_j=e^{i k_2}x_j, \quad j=1,2,3,4,\\
\text{Matching conditions at $\mathbf{V}$:}
\left\{\begin{aligned}
x_3&=\mathcal{E}x_1+ie^{-\varepsilon\pi}\mathcal{E}x_2,\\
x_4&=ie^{-\varepsilon\pi}\mathcal{E}x_1+\mathcal{E}x_2,
\end{aligned}\right.\quad
\text{at $\Tilde{\mathbf{V}}$: }
\left\{\begin{aligned}
y_3&=\Tilde{\mathcal{E}}y_1+ie^{-\Tilde\varepsilon\pi}\Tilde{\mathcal{E}}y_2,\\
y_4&=ie^{-\Tilde\varepsilon\pi}\Tilde{\mathcal{E}}y_1+\Tilde{\mathcal{E}}y_2,
\end{aligned}\right.\\
\text{Holonomy equations: }
x_4=y_3,\quad
w_2=e^{i\alpha_1}x_1,\quad
z_3=e^{i\alpha_2}w_4,\quad
y_1=e^{i\alpha_3}z_2.
\end{gather*}
After some algebra one arrives at a $2\times 2$ linear system,
\[
\begin{pmatrix}
\mathcal{E}-\Tilde{\mathcal{E}}e^{i(\alpha_1+\alpha_2-k_2)}
& ie^{-\varepsilon\pi}\mathcal{E}-ie^{-\Tilde\varepsilon\pi}\Tilde{\mathcal{E}}
e^{i( k_1+\alpha_2+\alpha_3)}\phantom{\bigg|}\\
ie^{-\varepsilon\pi}\mathcal{E}-ie^{-\Tilde\varepsilon\pi}\Tilde{\mathcal{E}}
e^{i(\alpha_1-k_1)} & \mathcal{E}-\Tilde{\mathcal{E}}e^{i(k_2+\alpha_3)}
\end{pmatrix}
\begin{pmatrix}x_1\\x_2\end{pmatrix}=0.
\]
The condition for the last system to have non-trivial solutions
comes from the vanishing of its determinant and has the form
\begin{multline}
                \label{eq-cose}
\cos\bigg[
\arg \Gamma\Big(\frac{1}{2}+i\varepsilon\Big)-
\arg \Gamma\Big(\frac{1}{2}+i\Tilde\varepsilon\Big)
+(\varepsilon-\Tilde\varepsilon)\log h
-\frac{\alpha_1+\alpha_2+\alpha_3}{2}
\bigg]\\
=\frac{1}{\sqrt{\big(1+e^{-2\varepsilon\pi}\big)\big(1+e^{-2\Tilde\varepsilon\pi}\big)}}
\bigg[
e^{-(\varepsilon+\Tilde\varepsilon)\pi}
\cos\Big(k_1-\frac{\alpha_1-\alpha_2-\alpha_3}{2}\Big)
+\cos\Big(k_2-\frac{\alpha_1+\alpha_2-\alpha_3}{2}\Big)
\bigg].
\end{multline}
Introducing again the weights $\beta_j$ of the edges $\gamma_j$, $j=1,2,3,4$,
by the rule
\begin{gather*}
\alpha_1=\beta_4+\beta_1,\quad \alpha_2=\beta_3-\beta_1,\quad\alpha_3=\beta_2-\beta_1,
\quad \arg\hol(\gamma_2+\gamma_3)=\beta_2+\beta_3,\\
\beta_j=\frac{B_j}{h}+\lambda J_j+\frac{\pi m_j}{2}+O(h),
\quad B_j=\int_{\gamma_j}p\ dx,\quad
m_j=\Ind\gamma_j,\quad j=1,2,3,4.
\end{gather*} we rewrite~\eqref{eq-cose} as
\begin{multline}
                \label{eq-cose2}
\cos\bigg[
\arg \Gamma\Big(\frac{1}{2}+i\varepsilon\Big)-
\arg \Gamma\Big(\frac{1}{2}+i\Tilde\varepsilon\Big)
+(\varepsilon-\Tilde\varepsilon)\log h
-\frac{\beta_1+\beta_2+\beta_3+\beta_4}{2}
\bigg]\\
=\frac{
e^{-(\varepsilon+\Tilde\varepsilon)\pi}
\cos\Big(k_1-\frac{(\beta_1+\beta_4)-(\beta_2+\beta_3)}{2}\Big)
+\cos\Big(k_2-\frac{(\beta_3+\beta_4)-(\beta_1+\beta_2)}{2}\Big)
}{\sqrt{\big(1+e^{-2\varepsilon\pi}\big)\big(1+e^{-2\Tilde\varepsilon\pi}\big)}}.
\end{multline}
Representing again $\lambda=\lambda_0+\mu$, $\mu=o(1)$, we come to the following
expression for $\mu$:
\begin{equation}
           \label{eq-disp3}
\begin{gathered}
\mu=\mu^\pm_n(k_1,k_2,\lambda_0,h)=\frac{w\Tilde w}{(w+\Tilde w)\log h}\\
{\times}\bigg\{
N(\lambda_0,h)\pm
\arccos\frac{e^{-\pi\lambda_0/w}\cos\big(k_1-\Delta_1(\lambda_0,h)\big)
+e^{-\pi\lambda_0/\Tilde
w}\cos\big(k_2-\Delta_2(\lambda_0,h)\big)}{\sqrt{(1+e^{-2\pi\lambda_0/w})(1+e^{-2\pi\lambda_0/\Tilde
w})}}
+2\pi n\bigg\}\\
{}\times\Big(1+O\big(\frac{1}{\log h}\big)\Big)+O(h),
\end{gathered}
\end{equation}
where
\begin{align*}
N(\lambda_0,h)&=\frac{B_1+B_2+B_3+B_4}{2h}
 +\frac{J_1+J_2+J_3+J_4}{2}\ \lambda_0+\frac{\pi(m_1+m_2+m_3+m_4)}{2}\\
&\quad {}-(\frac{1}{w}+\frac{1}{\Tilde w})\lambda_0\log h-\arg\Gamma\Big(\frac{1}{2}+\frac{i\lambda_0}{w}\Big)
+\arg\Gamma\Big(\frac{1}{2}-\frac{i\lambda_0}{\Tilde{w}}\Big),\\
\Delta_1(\lambda_0,h)&=\frac{(B_1+B_4)-(B_2+B_3)}{2h}+\lambda_0\frac{(J_1+J_4)-(J_2+J_3)}{2},\\
\Delta_2(\lambda_0,h)&=\frac{(B_3+B_4)-(B_1+B_2)}{2h}+\lambda_0\frac{(J_3+J_4)-(J_1+J_2)}{2}.
\end{align*}
The bandswidth $B(\lambda_0,h)$ and the gapwidth $G(\lambda_0,h)$
in a $o(h)$-neighborhood of the point $E+\lambda_0h$ admit a simple estimate
\begin{gather}
          \label{eq-b3}
B(\lambda_0,h)=\frac{2w \Tilde w h}{(w+\Tilde w)|\log h|}
\arcsin\frac{e^{-\pi\lambda_0/w}+e^{-\pi\lambda_0/\Tilde w}}
{\sqrt{(1+e^{-2\pi\lambda_0/w})(1+e^{-2\pi\lambda_0/\Tilde w})}}+o\Big(\frac{h}{\log
h}\Big),\\
          \label{eq-g3}
G(\lambda_0,h)=\frac{2w \Tilde wh}{(w+\Tilde w)|\log h|}
\arccos\frac{e^{-\pi\lambda_0/w}+e^{-\pi\lambda_0/\Tilde w}}
{\sqrt{(1+e^{-2\pi\lambda_0/w})(1+e^{-2\pi\lambda_0/\Tilde w})}}
+o\Big(\frac{h}{\log h}\Big),
\end{gather}
so that the bands clearly show an exponential decay with respect to $|\lambda_0|$.

\section{Discussion}\label{sec6}

In this section, we discuss in greater detail the influence of characteristics
of the Hamiltonian $H$ on the dispersion relation.

In the case of subsection~\ref{subsec41} the picture is quite simple.
The semiclassical dispersion relations depend on one of the quasimomenta only,
and the extremum points (with respect to this quasimomentum $k_2$)
is determined by the difference between the upper and the lower part
of the separatrix, $\gamma_1$ and $\gamma_2$; the width of the bands
and the gaps, which is given by Eqs.~\eqref{eq-b1} and ~\eqref{eq-g1} respectively,
depends on the determinant of the second derivatives at the critical point;
these formulas present a generalization
of a similar estimate for the periodic Sturm-Liouville problem
(see, for example, \cite[Sec.~10]{cdv-par}).

The example considered in subsection~\ref{subsec42} and figure~\ref{fig2}
differs
from the previous one. The position of the extremum point
with respect to $k_2$ is, like in the previous case, determined by
the relationship between the upper $\gamma_2+\gamma_4$
and the lower $\gamma_1+\gamma_3$ parts of the separatrix.
But the band- and the gapwidths, as can be seen from~\eqref{eq-b2},
\eqref{eq-g2a}, and~\eqref{eq-g2b}, crucially depend on $\Delta_1$;
the quantity $\Delta_1$
can be interpreted as a ``difference'' between the cycles $\gamma_1+\gamma_2$
and $\gamma_3+\gamma_4$. For example, if the areas of these two cycles
do not coincide, the ratio bandwidth/gapwidth has no limit for $h\to 0$. 
Examples with more complicated separatrices
will show a more curious picture.

\begin{figure}\centering
\begin{minipage}{30mm}\centering
\includegraphics[width=25mm]{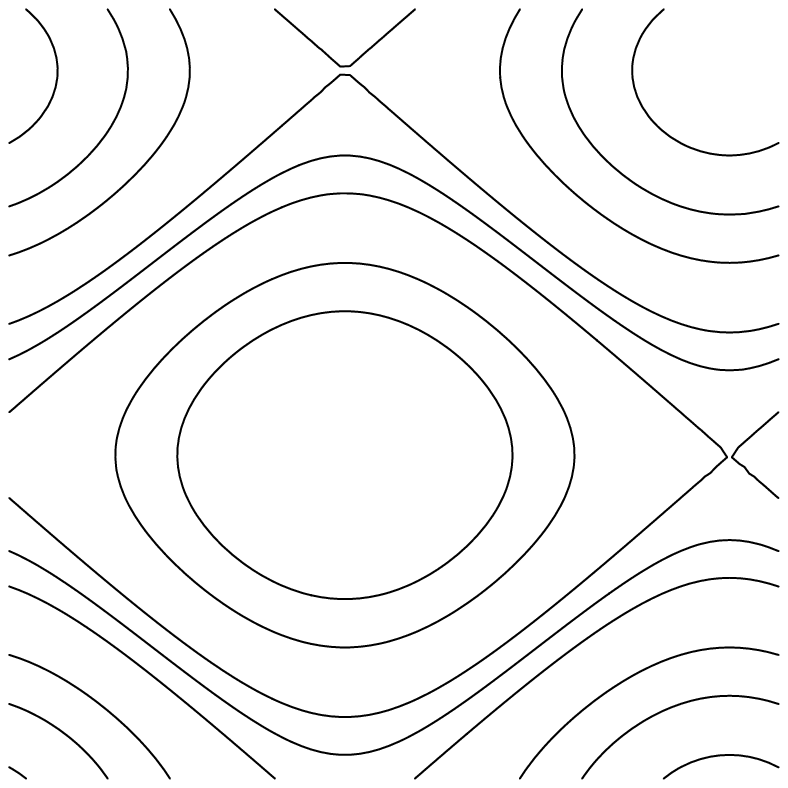}\\
$\lambda_0>0$
\end{minipage}\qquad
\begin{minipage}{30mm}\centering
\includegraphics[width=25mm]{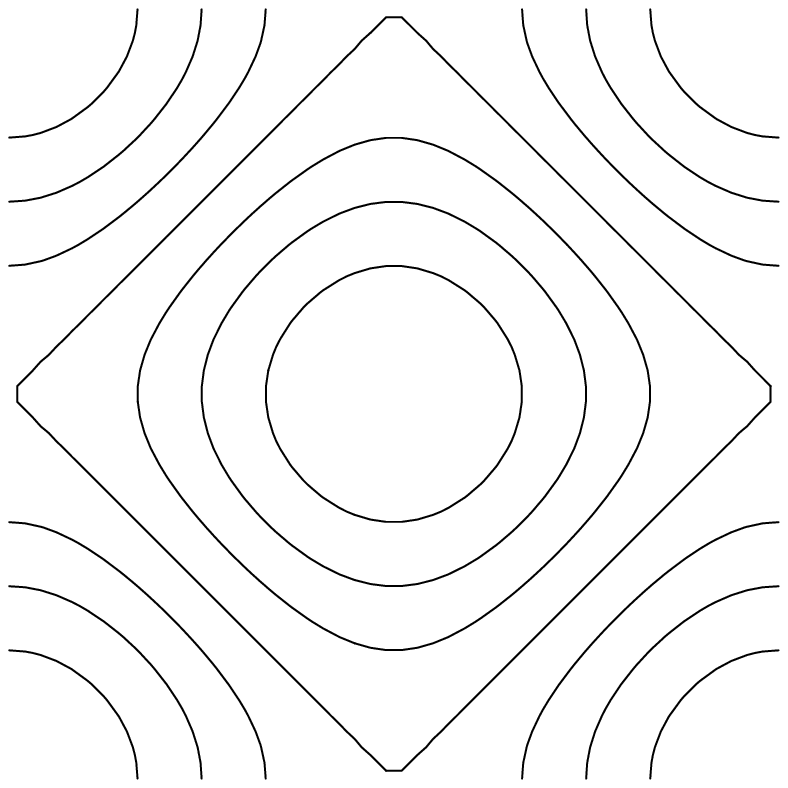}\\
$\lambda_0=0$
\end{minipage}\qquad
\begin{minipage}{30mm}\centering
\includegraphics[width=25mm]{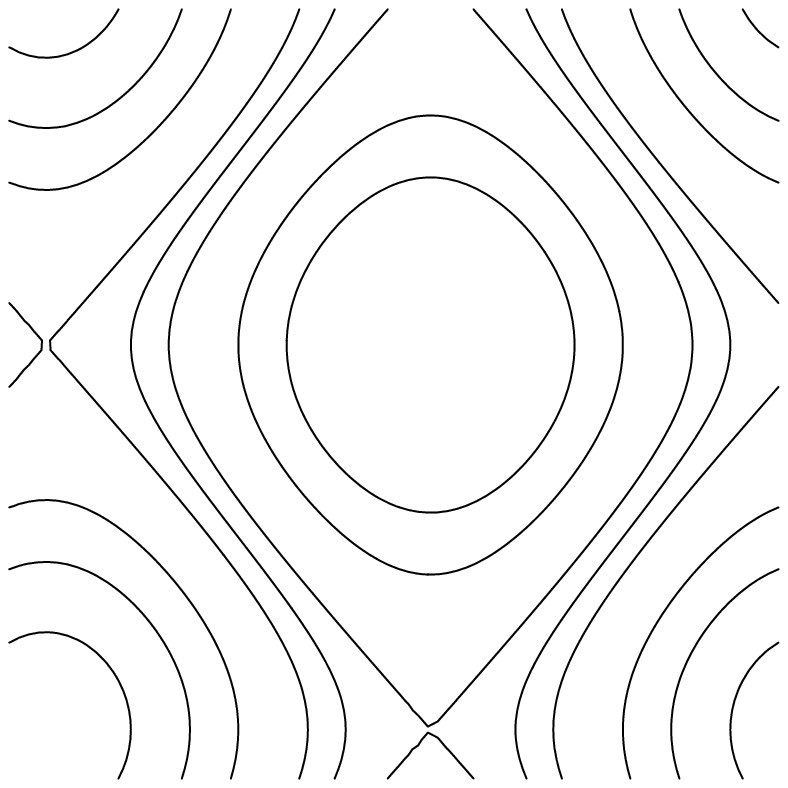}\\
$\lambda_0<0$
\end{minipage}
\caption{Levels curves of the dispersion relations lying near $E+\lambda_0h$}
           \label{fig-disp-ex}
\end{figure}

The example of section~\ref{sec5} shows a regular behavior
of the band and the gaps; the main term in the asymptotic
of their width, Eqs.~\eqref{eq-b3} and~\eqref{eq-g3},
like in the first example,
is determined by the second derivatives at the critical points only.
From the other side, this example is suitable for discussing
the form of the energy bands.
The dispersion relations lying in a neighborhood of $E+\lambda_0h$
have maxima at the points
\begin{equation}
              \label{eq-Kmax}
\mathbf{k}=\mathbf{k}_{\mathrm{max}}(\lambda_0,h):=\bigg(2\pi\times\Big\{\frac{\Delta_1(\lambda_0,h)}{2\pi}+\frac{1}{2}\Big\}-\pi,
2\pi\times\Big\{\frac{\Delta_2(\lambda_0,h)}{2\pi}+\frac{1}{2}\Big\}-\pi\bigg)
\end{equation}
(here $\{x\}$ denotes the fractional part of $x$).
In the generic situation, when all the holonomies $\beta_j$ are different,
these points depend crucially on $\lambda_0$, $B_j$, $I_j$,
and $h$, so that a small variation of them can change $\mathbf{k}_{\mathrm{max}}$
significantly.
Near the critical energy $E$ (i.e. for $\lambda_0=0$),
the quasimomenta have ``equal rights'', i.~e.
the coefficients before $\cos$-terms in~\eqref{eq-disp3}
are approximately equal to $1$.
In the non-symmetric case, when $w\ne\Tilde w$, the situation changes
for non-zero $\lambda_0$. So, if $w>\Tilde w$, the quasimomentum $k_1$ dominates
if $\lambda_0>0$ and $k_2$ dominates for $\lambda_0<0$, and,
at the same time the maxima of the dispersion relations moves according to~\eqref{eq-Kmax},
see illustration in figure~\ref{fig-disp-ex}.
This gives a rather rough (but at the same time generic)
impression about the dispersion relation structure near the critical point.

\section*{Appendix. The periodic Landau Hamiltonian with a strong magnetic field
and Harper-like operators}

In this appendix, we briefly describe the relationship between the periodic
Landau operator and Harper-like operators as it was established
and studied in~\cite{bdgp,bdp,bdp-tmf}.

The periodic Landau Hamiltonian has the form
\[
\Hat L:=\frac{1}{2}\Big(-ih\frac{\partial}{\partial x_1}+x_2\Big)^2
-\frac{h^2}{2} \frac{\partial^2}{\partial x_2}+\varepsilon v(x_1,x_2),
\]
where $v$ is a two-periodic function with periods
$\mathbf{a}$ and $\mathbf{b}$. If $\varepsilon=0$, 
the spectrum of $\Hat L$ consists of infinitely degenerate eigenvalues
$I_n=(n+1/2)h$, $n\in\mathbb{Z}_+$, called \emph{Landau levels}.
The presence of non-zero $\varepsilon$
leads to a broadening of these numbers into a certain sets called
\emph{Landau bands}. We are going to show, under assumption that
both $h$ and $\varepsilon$ are small, that the broadening
of each Landau level is described by a certain Harper-like operator.

The corresponding to $\Hat L$ classical Hamiltonian is
$L(p,x,\varepsilon)=(p_1+x_2)^2/2+p_2^2/2+\varepsilon v(x_1,x_2)$.
If $\varepsilon=0$, then $L$ defines an integrable system whose trajectories
on the $(x_1,x_2)$-plane are cyclotron orbits. For non-zero $\varepsilon$ the Hamiltonian
is non-integrable, but, for small $\varepsilon$, one can interpret
the classical dynamics as a cyclotron motion around a guiding center.
We introduce new canonical coordinates connected the motion of the center:
$p_1=-y_2$, $p_2=-q$, $x_1=q+y_1$, $x_2=p+y_2$,
considering $(p,y_1)$ as generalized momenta and $(q,y_2)$
as generalized positions ($p,q$ describe the motion around
the center with the coordinates $y_1,y_2$), then $H$ takes the form
$L=(p^2+q^2)/2+\varepsilon v(q+y_1,p+y_2)$.
Introduce the averaged Hamiltonian
\[
\Bar L(I,y_1,y_2,\varepsilon)=I+\frac{1}{2\pi}\oint
v(\sqrt{2I}\sin\varphi+y_1,\sqrt{2I}\cos\varphi+y_2) \,d\varphi
\equiv I+J_0(\sqrt{-2I\Delta_y})v(y_1,y_2),
\]
where $J_0$ is the Bessel function of order zero and
$\Delta_y=\partial^2/\partial y_1^2+\partial^2/\partial y_2^2$.
One can show that there exists a \emph{canonical} change of variables
$(p,y_1,q,y_2)=(\bar p,\bar y_1,\bar q,\bar y_2)+O(\varepsilon)$,
periodic in $y_1,y_2$ with periods $\mathbf{a}$ and $\mathbf{b}$,
such that
$L=\bar L\big(({\bar p}^2+{\bar q}^2)/2,\bar y_1,\bar
y_2,\varepsilon\big)+O(\varepsilon^2)$.
The averaging procedure can be iterated, so that one constructs
an averaged Hamiltonian
$\mathcal{L}=\mathcal{L}(J,Y_1,Y_2,\varepsilon)$
and a \emph{canonical} change of variables $(p,y_1,q,y_2)=
(P,Y_1,Q,Y_2)+O(\varepsilon)$, both periodic in $y_1,y_2$,
such that
$L=\mathcal{L}\big((P^2+Q^2)/2,Y_1,Y_2,\varepsilon\big)+O(\varepsilon^\infty)$.
Therefore, neglecting the last term and using the canonicity of all the transformations
one reduces the spectral problem for $\Hat L$ to that for
$\Hat{\mathcal{L}}$ obtained from $\mathcal{L}$ by the Weyl quantization
($P=-ih\partial/\partial Q$, $Y_1=-ih\partial/\partial Y_2$):
\begin{equation}
         \label{eq-calfi}
\Hat{\mathcal{L}}\Psi(Q,Y_2)=E\Psi(Q,Y_2).
\end{equation}
Clearly, $\Hat{\mathcal{L}}$ commutes with the harmonic oscillator
$\Hat I:=(-h^2\partial^2/\partial Q^2+Q^2)/2$, which means that the eigenfunction
$\Psi$ in~\eqref{eq-calfi} can be represented as
$\Psi(Q,Y_2)=\psi_n(Q)\Phi(Y_2)$, where $\psi_n$ is an eigenfunction
of $\Hat I$ with the eigenvalue $I_n$, and $\Phi$ must be an eigenfunction
of the operator $\Hat L_n$ obtained by quantizing the classical Hamiltonian
$L_n(Y_1,Y_2)=\mathcal{L}(I_n,Y_1,Y_2,\varepsilon)$ considering $Y_1$ as a momentum
and $Y_2$ as a position. All these Hamiltonians $L_n$ 
are periodic in $Y_1$ and $Y_2$, therefore, after a linear symplectic transformation
$\Hat L_n$ becomes a certain Harper-like operator which can be treated
as a Hamiltonian describing the broadening of Landau level under the presence of the electric
potential $v$.

\begin{figure}\centering
\begin{minipage}{45mm}\centering
\includegraphics[width=30mm,height=15mm]{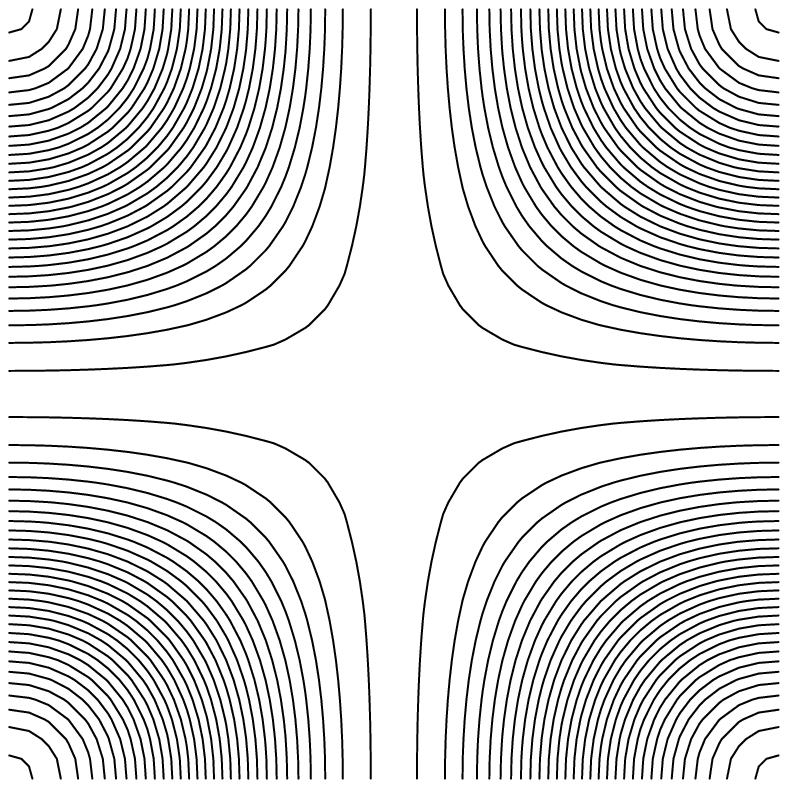}\\(a)
\end{minipage}
\begin{minipage}{45mm}\centering
\includegraphics[width=30mm,height=15mm]{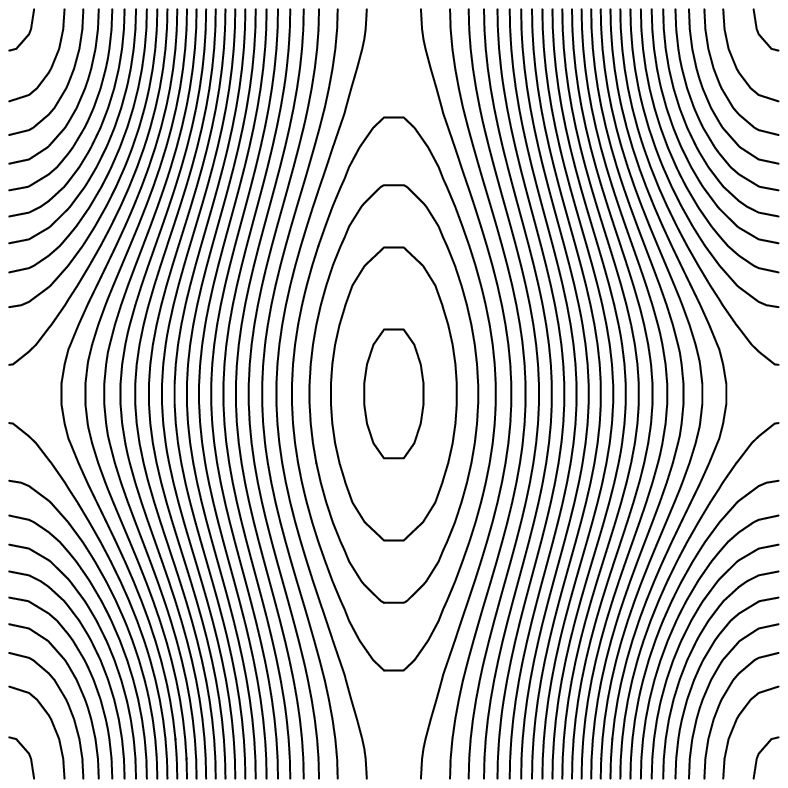}\\(b)
\end{minipage}
\begin{minipage}{45mm}\centering
\includegraphics[width=30mm,height=15mm]{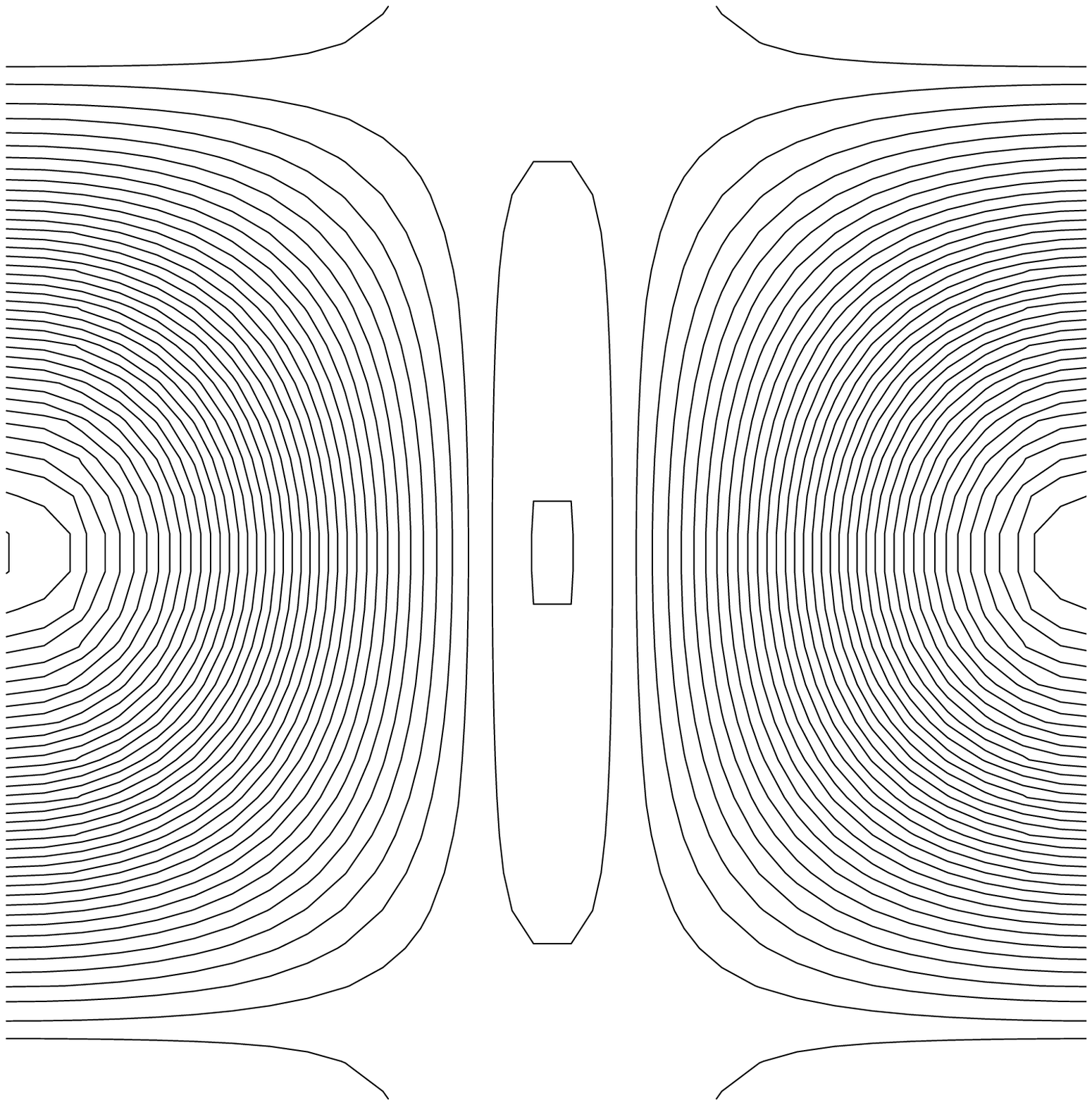}\\(c)
\end{minipage}
\caption{Level curves in the elementary cell $[0,2\pi)\times[0,\pi)$: (a) of the potential $v$,
(b) of the averaged Hamiltonian for $I\approx 0.5$,
(c) of the averaged Hamiltonian for $I\approx 1.5$}
\label{fig-ex}
\end{figure}

An essential point in our considerations is the dependence of $L_n$ on $n$:
one has
$L_n(Y_1,Y_2)=I_n+\varepsilon J_0(\sqrt{-2 I_n\Delta_Y})v(Y_1,Y_2)
+O(\varepsilon^2).$
In general, the topology of the trajectories of $L_n$ depends on
$n$, and a given potential $v$ can produce a number of operators
$\Hat L_n$ defining the structures of different Landau bands.
Considering a simple example $v=\cos^2 (x_1/2)\cos^2 x_2$
one arrives at $\mathcal{L}(I,Y_1,Y_2,\varepsilon)=I+\varepsilon\big(1+
J_0(\sqrt{2I})\cos x_1+J_0(\sqrt{8I})\cos 2x_2+
J_0(\sqrt{10I})\cos x_1 \cos 2x_2)\big)/4+O(\varepsilon^2)$. The level
curves of the potential $v$ and of the averaged Hamiltonian
are sketched in figure~\ref{fig-ex}; obviously, they are 
different, which implies differences in the structure of the corresponding
Landau bands.

\section*{Acknowledgments}

The author thanks J.~Br\"uning, S.~Dobrokhotov, and V.~Geyler for stimulating
discussions. The work was partially supported by the Deutsche
Forschungsgemeinschaft and INTAS.

\end{document}